\documentclass[pre,aps,amsmath,amssymb,amsfonts,floatfix,superscriptaddress,showpacs,twocolumn,nofootinbib,10pt]{revtex4-1}
\usepackage{graphicx, subfigure}
\usepackage{color}
\usepackage{bbm}
\usepackage[utf8]{inputenc}
\usepackage{dcolumn}
\usepackage{hyperref}
\usepackage{textcomp}
\usepackage[english]{babel}
\usepackage{natbib}
\usepackage{tikz}
\usepackage{mathtools}
\usepackage{amsmath}
\usepackage{dcolumn,overpic}% Align table columns on decimal point
\usepackage{bm}% bold math
\usepackage{tabularx}

\usepackage{comment}

\hypersetup{colorlinks=true,breaklinks,linkcolor=blue,urlcolor=blue,citecolor=blue}

\begin{document}
\title{A comparison between methods of analytical continuation for bosonic functions}

\author{J. Sch\"ott}
\affiliation{Dept.\ of Physics and Astronomy, Uppsala University, Box 516, SE-75120 Uppsala, Sweden}
\author{E. G. C. P. van Loon}
\affiliation{Institute of Molecules and Materials, Radboud University of Nijmegen, Heyendaalseweg 135, 6525 AJ Nijmegen, The Netherlands}
\author{I. L. M. Locht}
\affiliation{Dept.\ of Physics and Astronomy, Uppsala University, Box 516, SE-75120 Uppsala, Sweden}
\affiliation{Institute of Molecules and Materials, Radboud University of Nijmegen, Heyendaalseweg 135, 6525 AJ Nijmegen, The Netherlands}
\author{M. I. Katsnelson}
\affiliation{Institute of Molecules and Materials, Radboud University of Nijmegen, Heyendaalseweg 135, 6525 AJ Nijmegen, The Netherlands}
\author{I. Di Marco}
\affiliation{Dept.\ of Physics and Astronomy, Uppsala University, Box 516, SE-75120 Uppsala, Sweden}

\begin{abstract}
In this article we perform a critical assessment of different known methods for the analytical continuation of bosonic functions, namely the maximum entropy method, the non-negative least-square method, the non-negative Tikhonov method, the Pad\'e approximant method, and a stochastic sampling method. Three functions of different shape are investigated, corresponding to three physically relevant scenarios. They include a simple two-pole model function and two flavours of the non-interacting Hubbard model on a square lattice, i.e. a single-orbital metallic system and a two-orbitals insulating system. The effect of numerical noise in the input data on the analytical continuation is discussed in detail. Overall, the stochastic method by Mishchenko {\itshape{et al.}} [Phys. Rev. B \textbf{62}, 6317 (2000)] is shown to be the most reliable tool for input data whose numerical precision is not known. For high precision input data, this approach is slightly outperformed by the Pad\'e approximant method, which combines a good resolution power with a good numerical stability. Although none of the methods retrieves all features in the spectra in the presence of noise, our analysis provides a useful guideline for obtaining reliable information of the spectral function in cases of practical interest. 
\end{abstract}


\maketitle

\section{Introduction}
\label{sec:intro}

\begin{figure}[]
\includegraphics[width=0.8\columnwidth]{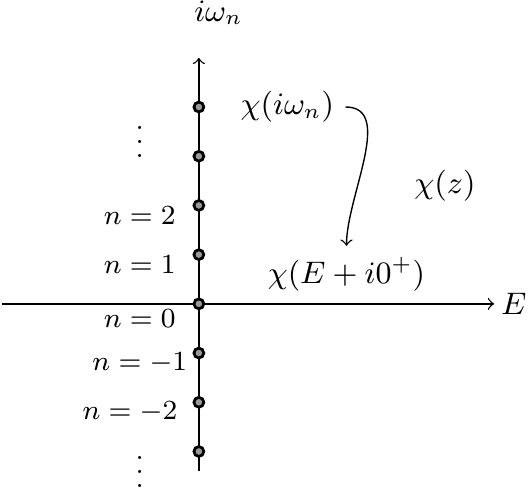}
\caption{Schematic illustration of the analytic continuation of a bosonic response function $\chi(z)$. The values $\chi(i\omega_n)$ at the Matsubara frequencies $i\omega_n$ are used to reconstruct the function $\chi(E+i\delta^+)$ just above the axis of real energies.}
\label{figtikz}
\end{figure}

Strongly correlated materials exhibit a wide range of exotic physical phenomena, ranging from magnetism to superconductivity~\cite{anisimov_book}. This rich physics is interesting for the scientific community since it has a high potential for future technological applications. In the last decades several theories have been developed to describe the electronic structure of strongly correlated materials with a good accuracy, e.g. the combination of density-functional theory (DFT) and dynamical mean-field theory (DMFT)~\cite{Metzner89,Georges96,Lichtenstein98,Kotliar06}. Practical calculations for systems at finite temperature are usually performed using the Green's function formalism for complex energies~\cite{NegeleOrland,Mahan}. In this approach determining physical observables requires an analytical continuation from complex energies (Matsubara frequencies) to real energies, as depicted in Fig.~\ref{figtikz}.

Traditionally, the analytical continuation of one-particle Green's functions has received the most attention, since it is related to the one-particle spectral function. The latter is not only of fundamental importance, but can also be probed (more or less) directly via various types of photoemission experiments. Over the last few years however, dynamical \emph{two-particle} quantities became more and more important~\cite{Boehnke12,Ayral12,Hansmann13,Ayral13,Huang14,vanLoon14,Hafermann14-2,Galler15,PhysRevLett.114.146401}. Significant examples include the electron energy loss spectrum (EELS)~\cite{Ayral12,vanLoon14} and dynamical susceptibilities~\cite{katanin10,Boehnke11,Galler15}.
Physically, the two-particle spectrum contains several interesting features, such as low-energy dispersive Goldstone modes, Landau damping of collective modes and sharp transitions between isolated energy levels. 
Even in a weakly interacting Fermi liquid, the two-particle spectrum presents some of these interesting features~\cite{Mahan}. From a more technical point of view, two-particle quantities also play an important role in the treatment of nonlocal interaction in DMFT-based approaches such as extended DMFT (EDMFT)~\cite{Si96,Kajueter96,Smith00,Chitra00,Chitra01} and its extension dual boson~\cite{Rubtsov12}, by describing the  feedback of collective excitations on the one-particle spectrum. 

Similarly to the one-particle Green's function, two-particle quantities are usually obtained on the Matsubara axis. Some physical information, such as the occurrence of charge-order transitions~\cite{Ayral13,vanLoon14-2,vanLoon15-2}, can be obtained directly from computational data on the Matsubara axis at zero-frequency. However, the full dynamical susceptibilities, as well as related observables such as the EELS and the plasmon spectrum, require an analytical continuation to real energies. It is easy to understand why finding a reliable method for the analytical continuation of two-particle functions has evolved from a niche problem to a necessity in computational many-body physics. 

The usual methods of analytical continuation of fermionic functions can, after some manipulation, also be used for bosonic functions~\cite{PhysRevB.82.165125}. It was previously shown that the maximum entropy method (MEM) is able to perform the analytical continuation of the optical conductivity with a good accuracy~\cite{PhysRevB.81.155107}. More recently,
in the context of EDMFT, Huang {\itshape{et al.}}~\cite{Huang14} focused on the retarted interaction, emphasizing that a modified version of MEM~\cite{Casula12} leads to a good analytical continuation. While these studies provide a very interesting and useful analysis, they do not cover the applicability of various methods to functions of different shapes and characters. So far, a more complete study of the analytical continuation problem for bosonic functions is lacking. In this work we intend to provide such an analysis, by considering five different methods of analytical continuation in three physically relevant scenarios. In addition to the MEM~\cite{PhysRevB.41.2380, Jarrell1996133,bryans,PhysRevB.44.6011,PhysRevE.81.056701,PhysRevE.82.026701,PhysRevB.81.155107,PhysRevB.82.165125}, we will also test the non-negative least-square (NNLS) method~\cite{NNLS}, the non-negative Tikhonov (NNT) method~\cite{L-curve}, the Pad\'e approximant (Pad\'e) method~\cite{PhysRevB.61.5147,Vidberg77}, and Mishchenko's stochastic sampling method~\cite{PhysRevB.62.6317,MishchenkoJulich}. 
Furthermore, we will investigate the effect of numerical noise on the quality of the continuation, in connection to the usage of quantum Monte Carlo techniques (see e.g. Ref.~\onlinecite{Gull11}).

This article is structured as follows; in Sec.~\ref{sec:greensfunction} we review bosonic Green's functions and their analytical and symmetry properties. 
In Sec.~\ref{sec:algorithms} a brief introduction to the various continuation schemes is given.  
Then, in Sec.~\ref{sec:testapproach}, we describe the different testing cases and highlight the reasons why they are interesting. 
In Sec.~\ref{sec:testresults}, we show and analyze the results of the tests.
Finally in Sec.~\ref{sec:testconclusions} we present the conclusions of our study, including a comparative summary of the performance of the continuation methods investigated.

\section{Bosonic Green's functions}
\label{sec:greensfunction}

The analytical continuation consists in obtaining a function $\chi(z)$ in the whole complex plane from a finite set of values. The focus in this work is on functions with bosonic symmetry that are known on a finite number of bosonic Matsubara frequencies $i\omega_n = 2n\pi T i$, where $T$ is the temperature and $n$ is an integer number. From this partial knowledge, we intend to extract the values of the function in the entire complex plane, and especially just above the real axis, for $z=E+i\delta$.

The analytical continuation problem can be formulated in terms of the spectral function $\rho(E)$ by means of the Hilbert transform
\begin{equation}
\chi(i\omega_n) = \int_{-\infty}^\infty dE \frac{1}{i\omega_n-E} \,\, \rho(E)\: .  
\label{hilberteq}
\end{equation}
For bosonic functions, the spectral function is an odd function, i.e. $\rho(E)=-\rho(-E)$, which allows us to simplify Eq.~\eqref{hilberteq} to
\begin{equation}
\chi(i\omega_n) = \int_0^\infty dE \frac{-2E}{\omega_n^2+E^2} \,\, \rho(E) \: . 
\label{hilbert4bosons}
\end{equation}
From this equation we find that for large Matsubara frequencies $\chi(i\omega_n)$ is asymptotically proportional to $\omega_n^{-2}$. Without the odd symmetry of $\rho$, one would obtain an asymptotic behavior as $\omega_n^{-1}$, like for the one-particle Green's function. Furthermore, $\chi(i\omega_n)$ is purely real and $\chi(i\omega_n)=\chi(-i\omega_n)$.
It follows from the analytical properties of $\chi(z)$ that $\chi(i \omega_n)$ decreases monotonously as a function of $i\omega_n$~ \cite{landau1980statistical}.
The correct symmetry can be enforced in the analytical continuation procedure either explicitly by using Eq.~\eqref{hilbert4bosons} or implicitly by including both positive and negative Matsubara frequencies as input points. This issue is studied in more detail in App.~\ref{sec:impose_mirrorsym}. 

\section{Continuation algorithms}
\label{sec:algorithms}
The analytical continuations are performed by means of five different methods. We use our own, in-house implementations for all algorithms except for Mishchenko's method. For Mishchenko's method we use an MPI-parallelized version of the original code presented in Ref.~\onlinecite{PhysRevB.62.6317}. Here, a brief introduction to these algorithms is given. 

\subsubsection{NNLS method}
Discretizing the integral in Eq.~\ref{hilbert4bosons} makes it possible to reformulate the problem as a system of linear equations
\begin{equation}
\chi(i\omega_n) = \sum_j  \underset{K_{n,j}}{\underbrace{f_j \frac{-2E_j}{\omega_n^2+E_j^2}}} \: \:  \rho_j \: ,
\label{eq:NNLS}
\end{equation}
where $f_j$ is a quadrature weight and $K_{n,j}$ is the matrix to invert. However,this approach would not work in practice, since obtaining $\rho(E)$, once  $\chi(i\omega_n)$ is given, is an ill-posed problem. The NNLS method  solves Eq.~\eqref{eq:NNLS} in a least-square sense and stabilises the solution by enforcing the known symmetry property $\rho(E)E\geq 0$. This means finding a non-negative solution to a least-square problem:
\begin{equation}
\underset{\bm{\rho \geq 0}}{\text{min}}  \:  \left \| \bm{\chi}-K \: \bm{ \rho } \right \|^2  \: .
\end{equation}
Here, ${\bm{\chi}}$ and ${\bm{\rho}}$ are vectors containing the values $\chi(i\omega_n)$ and $\rho_j$ on the discrete points of Equation \eqref{eq:NNLS}.
The NNLS problem can be solved iteratively as described extensively in Ref.~\onlinecite{NNLS}.

\subsubsection{NNT method}

For ill-posed problems, regularisations are commonly used. One of the most famous regularisation methods is the Tikhonov method~\cite{L-curve}. 
Applying the Tikhonov method to the problem of analytical continuation defines the NNT method as  
\begin{equation}
\underset{\bm{\rho \geq 0}}{\text{min}}  \:  \left \| \bm{\chi}-K \: \bm{  \rho } \right \|^2 +\alpha \left \| \bm{\rho} \right \|^2,
\label{eq:NNT}
\end{equation}
where $\alpha$ is a weight parameter that can be determined by the L-curve method~\cite{L-curve} which selects $\alpha$ corresponding to the smallest value of $\ln(\left \| \bm{\chi}-K \: \bm{ \rho}_\alpha  \right \|^2)+\ln(\left \| \bm{\rho}_\alpha \right \|^2)$.
Here $\bm{\rho}_\alpha$ denotes the solution of Eq.~\ref{eq:NNT} for a fixed $\alpha$.

\subsubsection{MEM}

Another famous regularisation method is the Maximum Entropy Method (MEM)~\cite{PhysRevB.41.2380, Jarrell1996133,bryans,PhysRevB.44.6011,PhysRevE.81.056701,PhysRevE.82.026701,PhysRevB.81.155107,PhysRevB.82.165125}.
This method maximises the Neumann entropy for the spectral function. Formulated as a minimisation problem, the equations become:
\begin{equation}
\underset{\bm{\rho \geq 0}}{\text{min}}  \:  \left \| \bm{\chi}-K \: \bm{ \rho } \right \|^2 +\alpha \left \| S[\bm{\rho}] \right \| \: ,
\end{equation}
where the entropy
\begin{align}
S[\rho]=&\int dE \rho(E) \ln \left(\frac{\rho(E) }{ m(E)}\right)  \notag \\
\approx&  \sum_j f_j \rho_j \ln \left(\frac{\rho_j}{m_j }\right) \: 
\end{align}
contains a default model $m(E)$, which can incorporate a priori knowledge about the spectral function.

\subsubsection{Pad\'e approximant method}
The Pad\'e approximant method~\cite{PhysRevB.61.5147,Vidberg77} is based on a fitting procedure. 
This method starts with a rational polynomial ansatz for $\chi(z)$ with unknown coefficients, and fits this ansatz to the known values at the Matsubara axis $\chi(i\omega_n)$. In this way one can reconstruct the function in the entire complex plane.
The Pad\'e method can be formulated as a matrix problem~\cite{PhysRevB.61.5147}, which improves its numerical accuracy. As recently proposed~\cite{Schott15}, one can enhance the stability of the continuations (especially in presence of Matsubara noise) by taking an average of several Pad\'e approximants, obtained by varying the number of Pad\'e coefficients and Matsubara points in the fitting procedure.

\subsubsection{Mishchenko's stochastic sampling method}
\label{text:mish}
Finally, we consider the stochastic sampling method proposed by Mishchenko {\itshape{et al.}}~\cite{PhysRevB.62.6317,MishchenkoJulich}. In this method the spectral function is approximated by a set $c$ of  rectangles $R$, i.e.: 
$$\rho_c(\omega)=\sum_{R \in c} R_{ \left \{ h,b,m \right \} }(\omega) \:.$$
The rectangles $R$ are defined by their height $h$, width $b$ and center position $m$. A stochastic algorithm updates the rectangles randomly and accepts the changes according to a Metropolis algorithm based on the difference between the known function $\chi(i \omega_n)$ and the corresponding function obtained from $\rho_c(\omega)$. Combining this update scheme with a deterministic minimization of the Matsubara difference further improves the performance of the method. An average over different independent Monte Carlo chains is taken to minimize the influence of noise. 
Since the algorithm uses a condition for the sum of the rectangle weights, we scale $\rho$ by introducing $\tilde{\rho}(E)= \frac{-2 }{ E \chi_0}\rho(E)$. Instead of finding $\rho$ by solving Eq.~\ref{hilbert4bosons}, we calculate $\tilde{\rho}$, which has to obey 
\begin{equation}
\tilde{\chi}(i\omega_n) = \frac{\chi(i\omega_n)}{\chi_0} = \int_0^\infty dE \frac{E^2}{\omega_n^2+E^2} \,\, \tilde{\rho}(E) \: .
\end{equation}
For $n=0$ we have the desired normalization condition $1=\int_0^\infty dE \,\, \tilde{\rho}(E)$. 
Once $\tilde{\rho}$ is found we scale back to obtain $\rho$. This rescaling can in principle be used for the other methods as well, with the exception of Pad\'e where the continuation is not done by solving Eq.~\ref{hilbert4bosons}.

\section{Model descriptions}
\label{sec:testapproach}
To investigate the quality of the analytical continuation provided by the different methods we focus on test functions that are known in the entire complex plane, including the real energy axis. The functions are first evaluated at the Matsubara energies and then analytically continued to the real axis. In an ideal situation, the solution obtained by analytical continuation coincide with the original exact values. However, in practice, some differences arise due to the limited precision and amount of input data.

The continuation becomes more difficult when we add numerical noise to the Matsubara data, which models what happens in real calculations as by, e.g., Monte Carlo methods.  
There, the noise typically scales inversely with the square root of the computational time. As a practical example, the numerical noise in dual boson calculations~\cite{vanLoon14-2} is often in the 0.1\% range.
The numerical noise is modelled by adding relative Gaussian noise to each Matsubara point, $\chi(i\omega_n) (1+\epsilon)$, where $\epsilon$ is sampled from a Gaussian distribution with zero mean and a standard deviation $\sigma$. 
We use the same $\sigma$ for all frequencies. In realistic Monte Carlo simulations, the noise level will usually depend on the Matsubara frequency $i\omega_n$ in a complicated way.

\begin{figure*}[]
\includegraphics[width=0.9\textwidth]{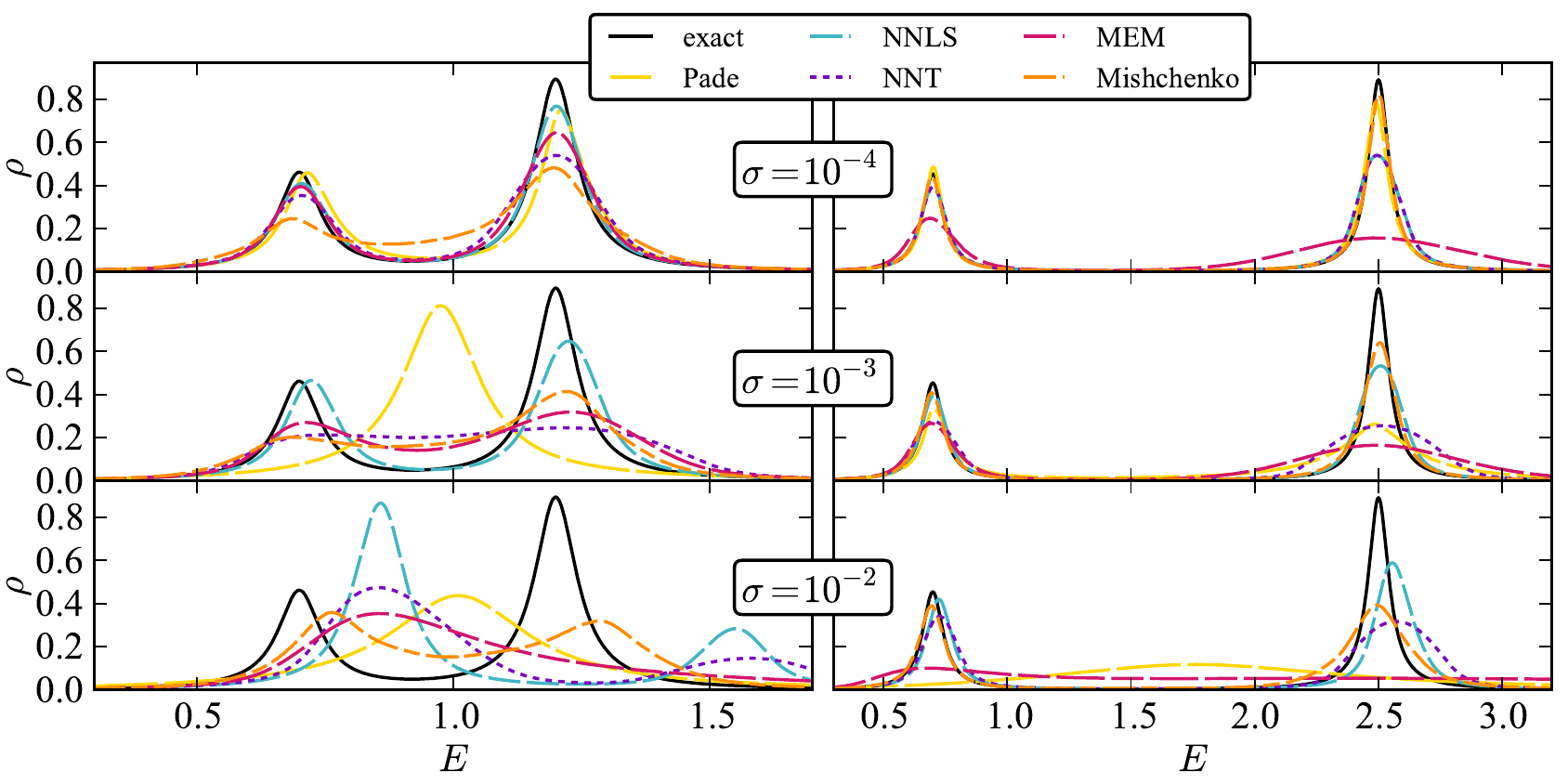}
\caption{Spectra for the two-pole test function. Left panel $a_1=0.1$, $a_2=0.335663$, $E_1=0.7$ and $E_2=1.2$, right panel $a_1=0.1$, $a_2=0.7$, $E_1=0.7$ and $E_2=2.5$, as defined in Eq.~\ref{eq:two-pole}.}
\label{fig:2pole:spectra}
\end{figure*}

It is important to individuate test functions that are realistic and cover bosonic functions of different character as one would encounter in practice. Naturally, the exact real energy solutions of many of the most interesting systems are not known, so we need to find realistic but tractable cases.
Here, we consider three scenarios: (i) a two-pole model, (ii) the momentum resolved random phase approximation (RPA) solution of the 2d doped non-interacting Hubbard model and (iii) that of a band insulator (with two identical bands separated by a gap). These three functions, which are explicitly described in Sec.~\ref{sec:testresults}, cover sharp features, a sharp low-energy zero sound mode (that gets broadened by Landau damping at larger momenta), high frequency modes, features far away from zero energy and a combination of sharp features with a broad base. 

All these functions are evaluated for Matsubara frequencies corresponding to an inverse temperature $\beta=50$ and at a distance above the real axis $\delta=0.05$. For all methods we used 100 Matsubara frequencies, except for Pad\'e where a spectral average is done by varying the number of Matsubara points between 50 to 100 in steps of 4. It is important to use a sufficiently large number of Matsubara frequencies. We study the dependence of the quality of the continuation on the number of Matsubara frequencies in Sec.~\ref{sec:two-polemod}. We also investigate the effect of including a few negative frequencies in Appendix~\ref{sec:impose_mirrorsym}. 

\section{Results}
\label{sec:testresults}
In this section we present a comparison between the different methods of analytical continuation when applied to three selected test functions.

\subsection{Two-pole model}
\label{sec:two-polemod}

The first function we address is a simple model function whose analytical structure is
\begin{equation}\label{eq:two-pole}
\chi(z) = \frac{a_1}{z^2-E_1^2}+\frac{a_2}{z^2-E_2^2} \: ,
\end{equation}
where the parameters have the following values $a_1=0.1$, $a_2=0.7$, $E_1=0.7$ and $E_2=2.5$. For $\delta \to 0^+$, the exact spectrum has two distinct peaks at energies $E_1$ and $E_2$. Therefore this function has similarities to the Gaussian two-peak structure used in Ref.~\onlinecite{Huang14}. 

To understand how the position of the high energy peak influences the quality of the attained spectra, we also perform tests with parameters $E_2=1.2$ and  $a_2=0.335663$. This choice preserves the height of both peaks in the two chosen setups. The two spectra, corresponding to $E_2=1.2$ and $E_2=2.5$, present different degrees of difficulty.  For $E_2=1.2$, the two peaks in the spectrum are close to each other and therefore it is difficult to resolve them as two separate peaks. For $E_2=2.5$, on the other hand, the second peak is far away from the first peak, but also far away from zero energy (the complex $i\omega_n$-axis). This makes it difficult to resolve the distant peak.

In Fig.~\ref{fig:2pole:spectra} the two exact spectra are reported, and compared to the analytical continuation obtained through the five methods for three different noise levels. It is known that the Pad\'e method is much more sensitive to the precision of the input data than the other methods. The various noise levels are therefore chosen to illustrate the critical input precision where the Pad\'e method starts to fail. These noise levels are realistic for Monte Carlo simulations~\cite{Gull11}.

As the top panels of Fig.~\ref{fig:2pole:spectra} show, for relatively low noise ($\sigma=10^{-4}$), all methods capture the two peaks and their positions correctly, for both sets of parameters. The Pad\'e method performs the best, if we also consider the height of the peaks. Conversely, the MEM has the most difficulties to resolve the distant peak at $E_2=2.5$. This picture changes drastically for intermediate noise levels ($\sigma=10^{-3}$), shown in the middle panels of Fig.~\ref{fig:2pole:spectra}. For $E_2=1.2$ (left panel), the Pad\'e method is not able to reproduce the two peaks, and finds only the one at low energy. The other methods also gain spectral weight in between the peaks, but are still able to resolve both peaks. For $E_2=2.5$ (right panel), results are significantly better for all methods. Interestingly, the MEM performs worse than the Pad\'e method. In situations where the input data is characterized by high noise ($\sigma=10^{-2}$), depicted in the bottom panels of Fig.~\ref{fig:2pole:spectra}, both the MEM and the Pad\'e method fail to find distinct peaks and merge them into one single broad peak. The other methods find distinct peaks, but locate them at the wrong energies, which is especially evident for $E_2=1.2$ (left panel). In any case, these methods are still able to describe the basic physics correctly, i.e. the presence of two well defined peaks. In general,  Mishchenko's method seems to offer the best results across various levels of noise. Even for the highest noise ($\sigma=10^{-2}$), Mishchenko's method gives an acceptable description of the exact spectra. Nevertheless one has to notice, if two peaks are close to each other, this approach tends to increase the spectral weight in between the peaks. 
As a matter of fact, this problem is not an intrinsic property of Mishchenko's algorithm, but is related to an insufficient number of local updates in the sampling. The dependence of Mishchenko's results with respect to the number of local updates is investigated in Appendix~\ref{app:conv_mish}. For the two-pole model, for small noise levels, Mishchenko's method can achieve results as good as Pad\'e, although at the cost of a much bigger computational effort.
 
Finally, from the comparison of the left and right panels of Fig.~\ref{fig:2pole:spectra}, one can conclude that partially overlapping peaks seem to be more difficult to describe than having one peak close and one peak far from zero energy.
This is not so trivial, since usually analytical continuation methods are supposed to perform well to describe (even complex) structures close to zero energy.

\subsubsection*{Number of Matsubara frequencies}
\begin{figure}[h!]
\includegraphics[width=0.9\columnwidth]{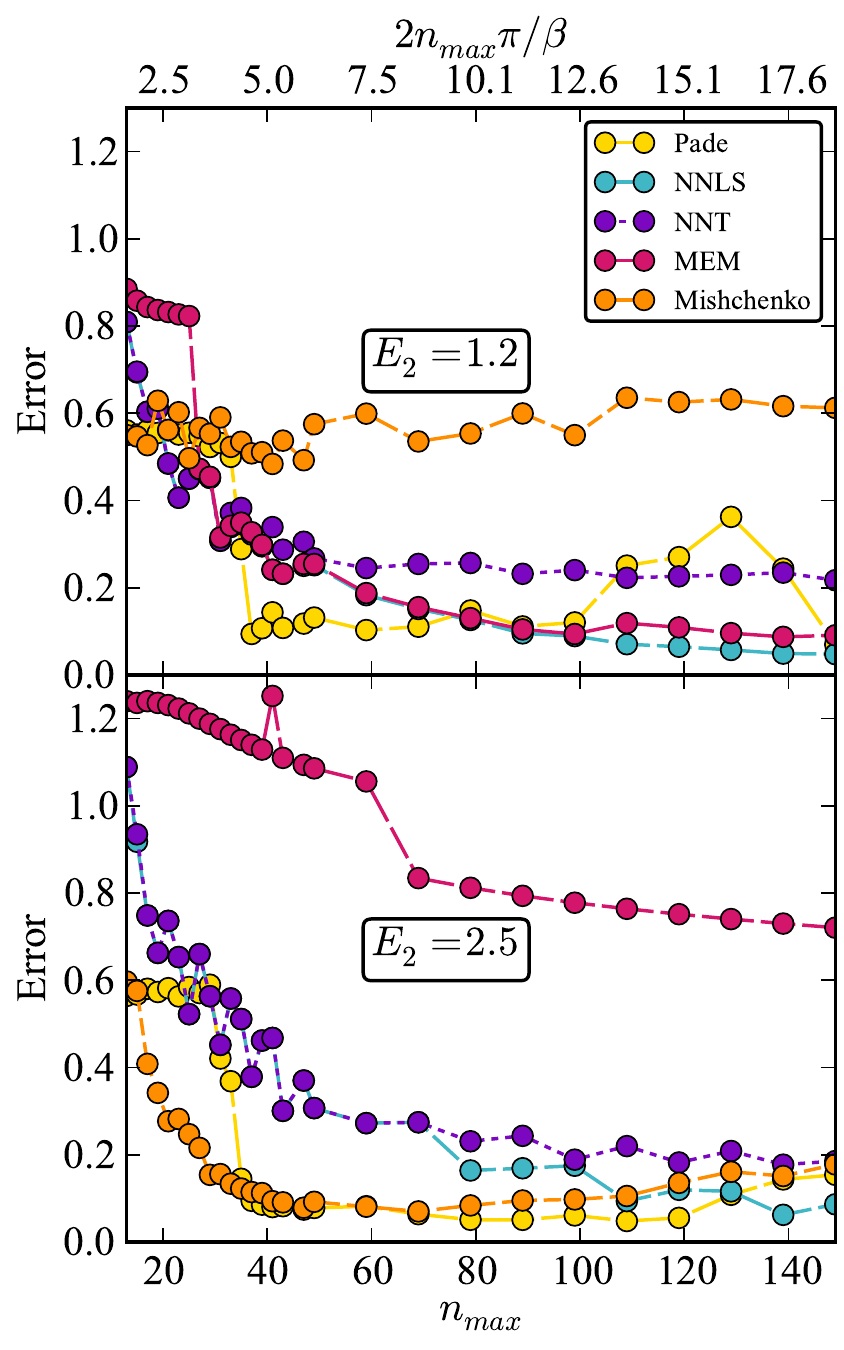}
\caption{Integrated real axis error for the two-pole model as a function of the highest Matsubara point index $n_\text{max}$ used. Upper panel $a_1=0.1$, $a_2=0.335663$, $E_1=0.7$ and $E_2=1.2$, lower panel $a_1=0.1$, $a_2=0.7$, $E_1=0.7$ and $E_2=2.5$, as defined in Eq.~\ref{eq:two-pole}. Matsubara noise level $\sigma=10^{-4}$ is used.}
\label{fig:2pole:error}
\end{figure}

It is also useful to illustrate how the quality of the analytical continuation depends on the number of input points, e.g., the number of Matsubara frequencies. The quality of the continuation can be measured through the difference between the obtained $\rho(E)$ and the exact function on the real axis $\rho_{\text{exact}}(E)$ in the following way
\begin{equation}
\text{Error} = \frac{\int_0^5 dE \left|\rho(E)-\rho_{\text{exact}}(E)\right|}{\int_0^5dE\left| \rho_{\text{exact}}(E) \right|} \: ,
\label{eq:error}
\end{equation}
where the upper bound of the integrals is chosen to be larger than the actual extension of the function. The error measure given by Eq.~\eqref{eq:error} for the two-pole model with noise $\sigma=10^{-4}$ is plotted in Fig.~\ref{fig:2pole:error} with respect to different number of the maximum Matsubara index, $n_{\text{max}}$. Note that for the Pad\'e method the notion of $n_{\text{max}}$ is a bit more complicated to define, since an average is done over several continuations with varying numbers of input points and Pad\'e coefficients~\cite{Schott15}. To make optimal usage of the averaging procedure, both the number of input points and Pad\'e coefficients take values between  $n_{\text{max}}-47$ and  $n_{\text{max}}+1$ in steps of 4, under the constraint of not having more coefficients than Matsubara points. If $n_{\text{max}}-47 < 4$, then 4 is used as the lower boundary instead.

In Fig.~\ref{fig:2pole:error}, the general trend is that the quality of the continuation initially improves with an increasing number of Matsubara points. At some point, the error stabilises. When the typical energy scale on the real axis ($E_2$) increases, this stabilisation point moves to higher Matsubara frequencies. However, there are small variations to this general picture for the different methods.

\begin{figure*}[t]
\includegraphics[width=\textwidth]{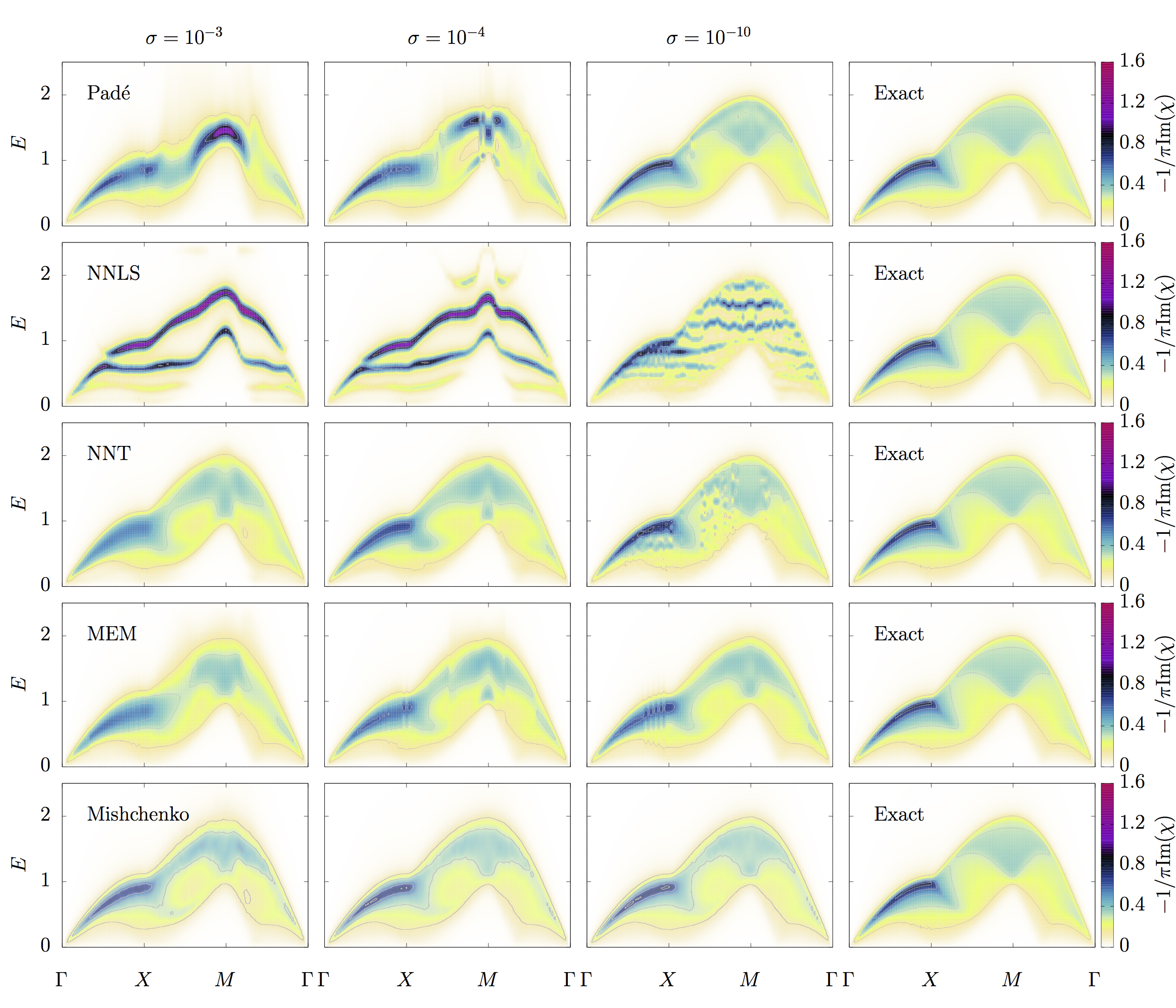}	
\caption{RPA spectrum ($-\frac{1}{\pi}\text{Im}[\chi(\omega+i\delta)]$) of the two-dimensional doped Hubbard model along the Brillouin zone path $\Gamma \to X \to M \to \Gamma$, obtained using Pad\'e, NNLS, NNT, MEM and Mishchenko's methods. The noise level varies among the values $\sigma=10^{-3}$, $\sigma=10^{-4}$ and $\sigma=10^{-10}$. All panels on the right hand side show the exact spectrum, for an easier comparison. The color map denotes the intensity of the spectra and contour lines are added at $-1/\pi\text{Im}(\chi)=0,0.15,0.30,0.45,0.60,0.75$. For the NNLS plots not all contour lines are displayed (see main text). The spectra for the $M$-point separately are also reported in Fig.~\ref{fig:RPA:noise:Pade_NNLS_NNT_MEM_2_inset}.}
\label{fig:RPA:noise:Pade_NNLS_NNT_MEM_2}
\end{figure*}

The error between the Pad\'e continuation and the exact function quickly drops at $n_{\text{max}} \approx 30$ and is very small until $n_{\text{max}} \approx 100$ (or 120 for $E_2=2.5$). When taking into account more Matsubara points, the error increases. We attribute this behavior mainly to the number of physical continuations in the averaging procedure. For small $n_{\text{max}}$, the total number of continuations is limited. For large $n_{\text{max}}$, the number of physical continuations decreases significantly. For this particular test, at least 30 to 40 physical continuations where needed, to optimally use the power of the averaging procedure. The continuations obtained with NNT and NNLS, on the other hand, continue improving when additional Matsubara frequencies are taken into account. However the improvement becomes smaller and smaller. The MEM error quickly drops at $2n_{\text{max}}\pi/\beta \approx 3 E_2$ and decreases only slightly for higher $n_{\text{max}}$. The error of the Mishchenko continuation stabilises at a relatively small number of Matsubara frequencies. For this method, we have used the same number of local updates (15000) for all continuations. The behaviour for more local updates is discussed in Appendix~\ref{app:conv_mish}.
It is conceivable that adding additional frequencies does give extra information, but also requires a longer runtime of the stochastic sampling.

\subsection{Doped Hubbard model}

\begin{figure}
\includegraphics[width=0.9\columnwidth]{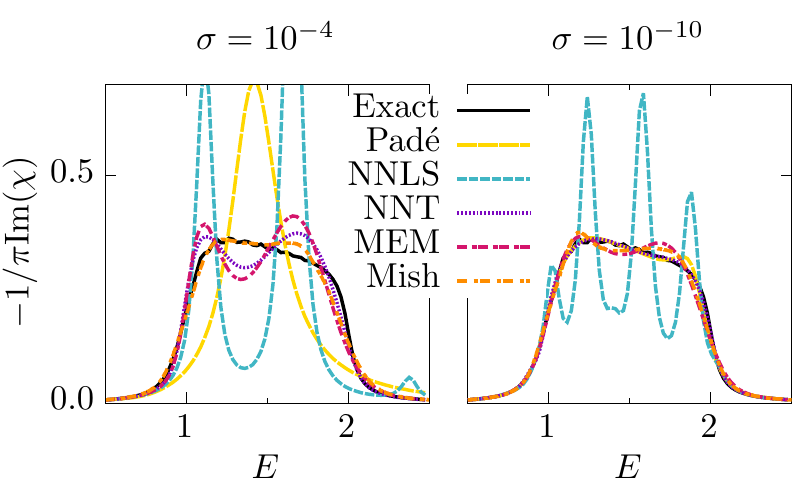}	
\caption{Profile at the $M$-point of the RPA spectrum of the doped Hubbard model reported in Fig.~\ref{fig:RPA:noise:Pade_NNLS_NNT_MEM_2}. Data for two noise levels $\sigma=10^{-4}$ and $\sigma=10^{-10}$ are shown. }
\label{fig:RPA:noise:Pade_NNLS_NNT_MEM_2_inset}
\end{figure}

For the second test, we use a metallic system, namely the two-dimensional non-interacting Hubbard model on a square-lattice with nearest-neighbor interaction. The correlation function depends on the Matsubara frequency and on the momentum $\mathbf{q}$. According to RPA it is given by~\cite{Pines66,Mahan}
\begin{align}
 \chi(i\omega_n,\mathbf{q}) = \sum_{\mathbf{p}} \frac{n_{F}(\epsilon_{\mathbf{p}})-n_{F}(\epsilon_{\mathbf{p}+\mathbf{q}})}{\epsilon_{\mathbf{p}} - \epsilon_{\mathbf{p}+\mathbf{q}} +i\omega_n},\label{eq:rpa}
\end{align}
where the sum denotes an average over the Brillouin zone, $n_F$ is the Fermi distribution and the dispersion relation is
\begin{align}
 \epsilon_{\mathbf{p}} = -2 t \left[ \cos p_x + \cos p_y \right].
\end{align}
The energy scale of this model is given by the half bandwidth $D=4t$, where  $t$ is the hopping parameter. Therefore, we fix the energy unit to $D=4t=1$. The chemical potential is set to $\mu=-0.5$, which corresponds approximately to a filling of 0.185 per spin flavor.
 
In Fig.~\ref{fig:RPA:noise:Pade_NNLS_NNT_MEM_2} the analytical continuation of the momentum resolved RPA susceptibility for all methods is reported. The panels in the first three columns correspond to data for various levels of noise $\sigma$. The panels in the last column illustrate the exact spectrum on the real axis, repeated in each row for an easier comparison.

At small wavevectors (close to $\Gamma$), the susceptibility exhibits a single low-energy mode, the zero sound mode. This mode consists of low energy excitations close to the Fermi surface. Further away from $\Gamma$, Landau damping broadens this zero sound mode. All methods correctly capture the zero sound mode. However, NNLS shows an unphysical splitting of this mode into two branches roughly halfway from $\Gamma \rightarrow X$. The tendency of NNLS to introduce spurious peaks is even more pronounced when going from $X \rightarrow M$, and in fact it is not even possible to illustrate all countour lines in Fig.~\ref{fig:RPA:noise:Pade_NNLS_NNT_MEM_2} without compromising its readibility. The complex spectral evolution in the path $X \rightarrow M$ seems indeed rather difficult to describe for all methods, especially for high noise. At the $M$-point the exact spectrum acquires a seemingly simpler structure, consisting of a relatively flat bump between $E=1$ and $E=2$. This point is analysed more in detail in Fig.~\ref{fig:RPA:noise:Pade_NNLS_NNT_MEM_2_inset}. It shows that it is not so easy to capture the broad flat feature instead of two separate peaks. Similarly to the previous test function, Pad\'e correctly reproduces the spectrum at low noise, but gives a single sharp peak instead of a broad flat mode when $\sigma$ is increased. NNLS performs very poorly here, with a spectrum of a few sharp peaks even at low noise $\sigma=10^{-10}$. NNT does a lot better, compared to NNLS; the regularisation flattens the spectrum, although an unphysical local minimum remains at $\sigma=10^{-4}$. Similarly, MEM also captures the general shape of the spectrum, but again an unphysical local minimum forms between the main peaks. Mishchenko's method is the only one that correctly reproduces the very flat spectrum between $E=1$ and $E=2$ even for high noise. These feature is consistent with the behavior observed in the top left panel of Fig.~\ref{fig:2pole:spectra}.

In general Mishchenko's method performs the best across various levels of noise, while NNLS leads to the poorest continuation for this function. As expected, the quality of the continuation obtained with Pad\'e worsens quickly when $\sigma$ is increased. Interestingly, NNT seems to perform slightly better for higher noise. For very low noise, the features are sharper and some spurious wiggles occur. These properties influence strongly the applicability of NNT and Pad\'e to realistic problems where the exact function is unknown. The MEM is rather robust to external noise, but for high precision data its quality seems inferior to both Mishchenko's method and Pad\'e.

\subsection{Band gap model}
\label{subsec:Band_gap_model}

\begin{figure*}
\includegraphics[width=0.95\textwidth]{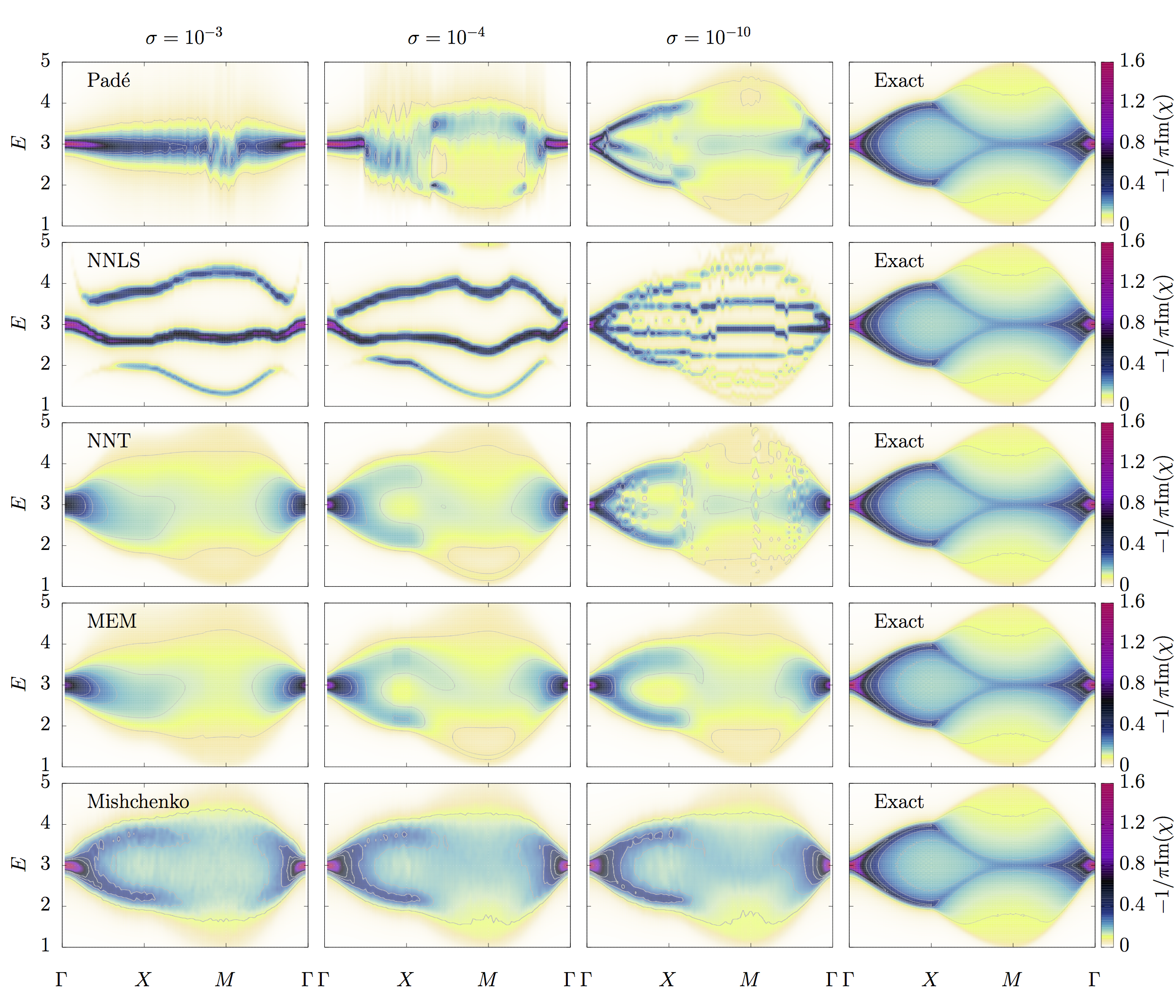}	
\caption{RPA spectrum of the band gap model ($-\frac{1}{\pi}\text{Im}[\chi(\omega+i\delta)]$) along the Brillouin zone path $\Gamma \to X \to M \to \Gamma$, obtained using Pad\'e, NNLS, NNT, MEM and Mishchenko's method. The noise level varies among the values $\sigma=10^{-3}$, $\sigma=10^{-4}$ and $\sigma=10^{-10}$. All right panels show the exact spectrum. The color map denotes the intensity of the spectra and contour lines are added at $-1/\pi\text{Im}(\chi)=0,0.1,0.2,0.3,0.4,0.5, 0.6$. The spectra at the $M$-point and at $\frac{1}{2}(\Gamma \rightarrow X)$ poit are reported separately in Fig.~\ref{fig:bandgap:noise:Pade_NNLS_NNT_MEM_inset}.}
\label{fig:bandgap:noise:Pade_NNLS_NNT_MEM}
\end{figure*}

\begin{figure}
     \subfigure[ Profile at $\frac{1}{2}(\Gamma \rightarrow X)$-point \label{subfig:GX}]{
         \includegraphics[width=\columnwidth]{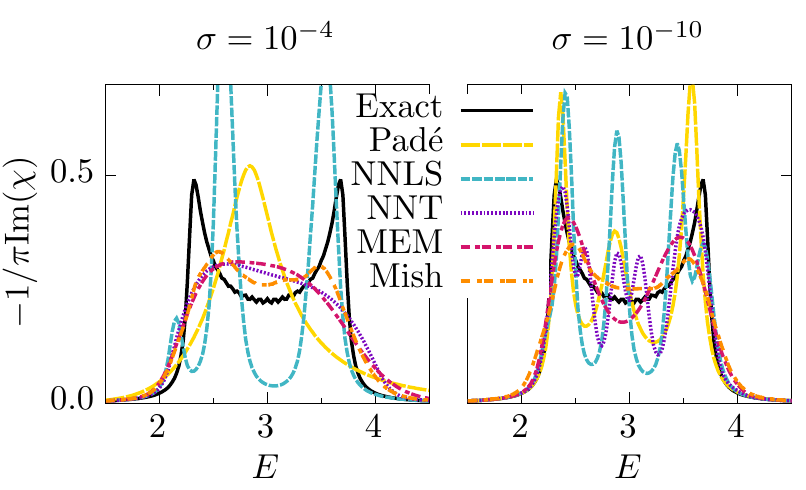}}
         %\hspace{1cm}
     \subfigure[ Profile at $M$-point \label{subfig:M}]{
         \includegraphics[width=\columnwidth]{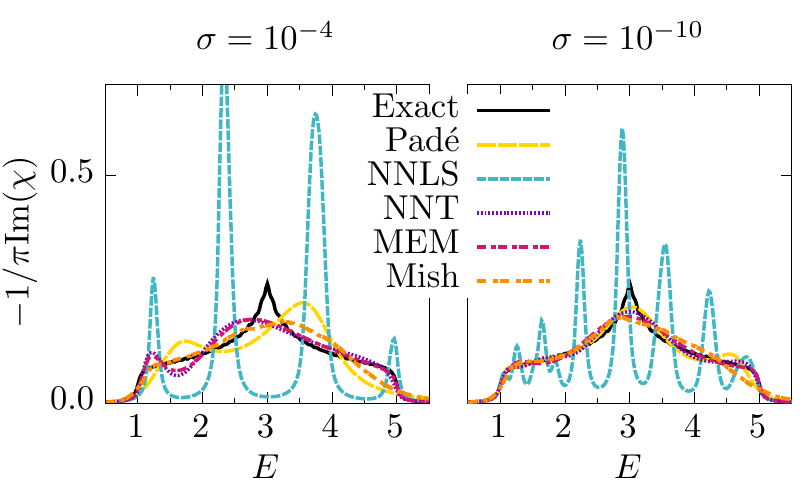}}
\caption{Profile at the $\frac{1}{2}(\Gamma \rightarrow X)$ and $M$-point of the RPA spectrum of the band gap model in Fig.~\ref{fig:bandgap:noise:Pade_NNLS_NNT_MEM} for two noise levels $\sigma=10^{-4}$ and $\sigma=10^{-10}$. }
\label{fig:bandgap:noise:Pade_NNLS_NNT_MEM_inset}
\end{figure}

The final model function is useful to investigate realistic features that are not close to zero energy. We again use a non-interacting Hubbard model on a square lattice with nearest-neighbor hopping. Differently from the previous test, we now consider two non-degenerate bands. Again, we take $4t=1$ for the half bandwidth. The two bands are shifted with respect to each other by an energy $E_\text{shift}=3$. Since the half bandwidth of both bands is 1, there is an indirect gap of $E_\text{gap}=1$. 
In Equation \eqref{eq:rpa}, this would correspond to two bands with shifted dispersions $\epsilon_\mathbf{p}$ and $\epsilon'_\mathbf{p}=\epsilon_{\mathbf{p}}+3$. We consider both excitations within one band as well as between the two bands.
The chemical potential $\mu=1.7$ lies in the gap between the two bands. This model could correspond to the presence of a magnetic field that splits the dispersion of the electrons with spins up and down. The electronic states with spin down have energy $E_\downarrow\in [-1,1]$, i.e. they are all below the Fermi energy. The electronic states with spin up, on the other hand, have energy $E_\uparrow\in [2,4]$. i.e. they are all above the Fermi energy. Excitations between the two bands have energies between $E=1$ and $E=5$. 

In Fig.~\ref{fig:bandgap:noise:Pade_NNLS_NNT_MEM} the analytical continuation of the momentum resolved RPA susceptibility for all methods is reported. As above, the panels in the first three columns correspond to data for various level of noise $\sigma$, while the panels in the last column illustrate the exact spectrum on the real axis.
At the $\Gamma$-point, the exact spectrum shows the presence of a single peak at $E=3$. This essential feature is captured by all methods and for all levels of noise, although the MEM and the NNT method exhibit the usual tendency to broaden the peak. Going from $\Gamma$ to $X$, the single peak splits into two distinct peaks, with some spectral weight remaining in the intermediate region. To illustrate this, Fig.~\ref{subfig:GX} shows the spectrum half way between $\Gamma$ and $X$. At $\sigma=10^{-4}$ MEM, NNT and Mishchenko's method all capture the widening of the peak as we move away from the $\Gamma$-point. However only Mishchenko's method captures the gradual splitting into two peaks at this point. MEM and NNT find the splitting closer to $X$. For smaller noise, i.e. for $\sigma=10^{-10}$, the MEM is also capable of resolving the two-peak structure at $\frac{1}{2} (\Gamma \rightarrow X)$. NNLS, NNT and  Pad\'e fail to offer an acceptable description of the spectrum at this point, since they also introduce one or more spurious peaks in between the two physical peaks. 

At the $M$ point, the exact spectrum has a sharp peak at $E=3$ with a very broad base. All methods, except for NNLS,  give the correct width, but are not very accurate on the symmetric character of the spectrum. This improves for lower noise levels, but the sharp peak in the middle stays slightly broadened, as can be seen in Fig.~\ref{subfig:M}. These data show that Mishchenko's method is the only one to offer a fully satisfactory description of the spectrum for $\sigma=10^{-4}$. It is also worth noticing that although the Pad\'e method does not fail too badly at $M$, the general trend for $\sigma=10^{-4}$ observed in Fig.~\ref{fig:bandgap:noise:Pade_NNLS_NNT_MEM} is rather negative and violates symmetry constraints. For even smaller noise, $\sigma=10^{-10}$, all methods but NNLS give a very good description of the spectrum at the $M$-point. 

In summary, for this model function, Mishchenko's method is stable with respect to input noise and clearly performs best of all methods for $\sigma=10^{-3}$. However, great care has to be taken concerning the number of updates in the sampling chain, as is investigated in Appendix~\ref{app:conv_mish}. The Pad\'e method performs quite well for high precision data but fails badly when noise is present. 
Finally, for intermediate levels of noise, MEM and NNT also lead to reasonable results, although Mishchenko's method is still superior.

\subsection{Computational time}

As a final note, we want to make a brief comment on the computational effort required to perform the analytical continuation of these functions.
Momentum resolved calculations, such as the RPA spectra for the Hubbard model shown above, typically involve linear dimensions of approximately 100 sites. Even considering a minimal effort, analytical continuation is required on a path through the high symmetry points of the lattice, involving about a hundred analytical continuations. Additionally, the analytical continuation should be done several times with different continuation parameters to ascertain its stability. Therefore, the computational time required to perform a single continuation becomes a factor of practical importance. 

It is not simple to compare the fundamental computational effort required by the five methods investigated here, due that the employed codes have different degrees of optimization and parallelization. Nevertheless, we can still provide some estimates. The fastest methods are undoubtedly NNLS and NNT, which require less than a second per continuation. The MEM and Pad\'e are slower, and usually require from several to a hundred seconds per continuation, depending on the parameters used.  Mishchenko's method is by far the most demanding method among those tested here. To obtain properly converged results one has to perform the stochastic sampling for a time ranging from a few hours to weeks. The effective time can be significantly reduced by using a parallelized code, like we do in the present work. However, parallelization is easy to implement only over global updates in the Monte Carlo chain, while local updates have still to be handled serially. This limits the scalability of the code, and therefore the reduction in computational time, to about hundred CPUs. Even in this case, we must notice that it is not always possible to dedicate such a large amount of resources to the analytical continuation problem. This sets a limit on the applicability of Mishchenko's method in current research. We are currently exploring an alternative formulation of Mishchenko's method based on a graphical processing unit (GPU)~\cite{GPU}. 

\section{Conclusions}
\label{sec:testconclusions}

\begin{table*}[]
\caption{Summary of the test performance.}
\label{tab:summary}
\begin{center}
\begin{tabular}{p{2.1cm}p{0.8cm} | p{2.5cm}p{2.8cm}p{2.9cm}p{2.8cm}p{2.7cm}}
				&$\sigma$&Pad\'e		&NNLS		&NNT				&MEM  & Mishchenko\\
\hline
% Two pole, left panel
2 close \newline poles &$10^{-2}$	&1 peak	&Broadened,  \newline positions wrong 	&Broadened,  \newline positions wrong	&1 peak & Broadened,  \newline positions wrong\\
 &$10^{-4}$	&Perfect  		&Good 	&Good 	&Good & Good\\[0.3cm]
% &&&&&&\\
% Two pole, right panel
2 separated &$10^{-2}$& 1 peak & Good & Good, broadened & 1 peak & Good, broadened \\
poles		&$10^{-4}$& Perfect & Good & Good & Good, broadened & Perfect \\[0.3cm]
%	&&&&&&\\
% Gamma-X points of doped Hubbard model
1 sharp 	&$10^{-3}$&Good	&Spurious features 	&Broadened	&Broadened & Good\\
feature	&$10^{-10}$&Excellent 	&Spurious features &Broadened, unstable	&Broadened & Good\\[0.3cm]
%	&&&&&&\\
% M point of doped Hubbard model
Broad, flat \newline feature	&$10^{-3}$&Too sharp	&2 sharp peaks	&Good, \newline unphysical  minima	&Good, \newline unphysical  minima	 & Good\\
&$10^{-10}$&Good	&Spurious features&Good	&Good & Good\\[0.3cm]
%&&&&&&\\
% X point of band gap model
2 peaks in \newline broad plateau	&$10^{-3}$&1 peak &Wrong positions \newline+additional peak &1 broad peak&1 broad peak & Good, broadened\\
       &$10^{-10}$&Additional peak  &Spurious peaks &Spurious peaks &Good, broadened &Good, broadened \\[0.3cm]
%&&&&&&\\
%M point of band gap model
1 peak in  \newline broad plateau 	&$10^{-3}$& 1 peak &Spurious peaks & Plateau &     Broad peak  \newline+plateau &Chopped triangle \\
      &$10^{-10}$& Good &Spurious peaks & Good &      Good &Good \\
\end{tabular}
\end{center}
\end{table*}

We have presented results for the analytical continuation of bosonic functions by five different methods and using three realistic test cases, corresponding to different scenarios of physical relevance for strongly correlated systems. An overview of the major features emphasized in the previous section is given in Table~\ref{tab:summary}. The first conclusion of our work is that, at the moment, for low input accuracy, none of the existing methods is good enough to be able to continue all different types of two-particle quantities from the Matsubara frequencies to real energies. Nevertheless, a combination of several methods can give enough information to reconstruct the physical picture behind the analytical continuation problem. Table~\ref{tab:summary} can offer useful guidelines to perform this type of analysis.

Conversely, for high precision input data, our results are encouraging. Among all methods, the Pad\'e averaging scheme~\cite{Schott15} is the one that performs best when the numerical noise is absent.
For complicated spectra, such as that of the band gap model, even Pad\'e cannot completely resolve the spectrum for $\sigma = 10^{-10}$, and it is indeed not even clear that it performs better than Mishchenko's method, due to small violations of the symmetry characterizing the exact spectrum. In general, Mishchenko's method seems to be the most robust approach across various levels of noise. As a matter of fact, its results do not seem very affected by the different noise levels for all the tested functions. 
The accuracy of Mishchenko's method is also good, although care is needed when choosing the numerical parameters of the stochastic sampling. For optimal parameter values, the accuracy of Mishchenko's method is definitely good.

Overall, a single sharp peak close to zero energy is resolved by all methods even for a relatively high noise. 
The MEM, NNT method and Mishchenko's method are able to resolve broad features. However, these methods also have the tendency to smear several pronounced peaks, sometimes into a single broad feature. In Mishchenko's method, this problem is related to the number of local updates performed in the sampling. As shown in Appendix~\ref{app:conv_mish}, increasing the number of local updates leads to more peaked spectra, but those are not always in better agreement with the exact results.
In general, the tendency to smear pronounced features into broad peaks and, at the same time, to create spurious peaks makes it difficult to know if a calculated broad . But on the other hand, sometimes spurious structure is created. This makes it difficult to know if a calculated broad spectral feature really is a broad feature or a combination of several peaks. As an example of this, the MEM continuation of the two-pole function for $E_2=1.2$ at $\sigma=10^{-4}$ (Fig.~\ref{fig:2pole:spectra}) looks similar to the MEM continuation of the two-dimensional doped Hubbard model at the $M$-point for the same noise level (Fig.~\ref{fig:RPA:noise:Pade_NNLS_NNT_MEM_2_inset}), although the exact functions are very different.

The very broad modes that appear, e.g., around the $M$ point of the doped two-dimensional Hubbard model, are difficult to resolve correctly. Functions with two distinct poles, on the other hand, are much easier. NNLS, in particular, performs well for the two-pole model whereas it performs very poorly for more complicated spectra. This shows the importance of varied, realistic test functions for assessing the potential of each continuation method.

Finally, we compare our work to previous studies. 
Two previous works~\cite{Huang14,PhysRevB.82.165125} found that Pad\'e generally performs poorer than the other methods. Huang \textit{et al.}~\cite{Huang14} found that MEM outperforms Pad\'e for a spectrum generated from a combination of two Gaussian distributions. In particular, their results show that the Pad\'e method sometimes merges two peaks, just as we have seen in the two-pole spectra.
In Ref.~\cite{PhysRevB.82.165125}, the authors test several methods of analytical continuation on the optical conductivity for relatively high noise levels ($\sigma=10^{-2}$ to $10^{-3}$). They find that all methods (Pad\'e, Singular Value Decomposition, sampling and MEM) perform similarly, but that the Pad\'e method generally gives slightly less accurate results and sometimes finds unphysical continuations. However, in both studies, the averaging scheme for Pad\'e~\cite{Schott15} was not used. This averaging scheme solves the problem of unphysical continuations and should improve the overall performance of the Pad\'e method. We would also like to stress two more fundamental conclusions of this study. First, we want to emphasize the importance of varied, realistic test functions in the assessment of continuation algorithms. The dependence of the performance of each method on the type of function investigated is a clear indication that no analysis can be meaningful without exploring a proper number of test cases.
Second, we want to stress that our findings fully support what was already claimed in  Ref.~\cite{PhysRevB.82.165125} , namely, despite no universal tool currently available, a better insight into the continuation problem can be obtained by using a combined approach of several methods.

\section{\label{sec:acknowledgments}Acknowledgments}
This work was sponsored by the Swedish Research Council (VR), the Swedish strategic research programme eSSENCE and the Knut and Alice Wallenberg foundation (KAW). E.G.C.P.v.L. and M.I.K acknowledge support from ERC Advanced Grant 338957 FEMTO/NANO. The authors acknowledge the computational resources provided by the Swedish National Infrastructure for Computing (SNIC) at the PDC Center for High Performance Computing at the KTH Royal Institute of Technology, at Uppsala Multidisciplinary Center for Advanced Computational Science (UPPMAX) and at Chalmers Centre for Computational Science and Engineering (C3SE).

%\clearpage
\appendix

\begin{figure}
	\subfigure[ The two-pole function in the presence of input noise with $\sigma=10^{-4}$\label{subfig:Mconv-2p}]{
		\includegraphics[]{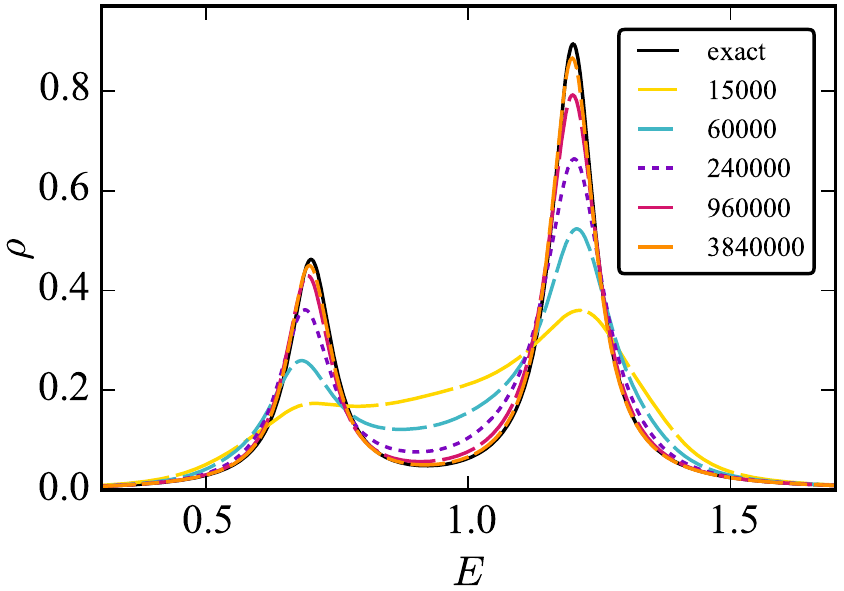}
	}
	
	\vspace{0.7cm}
	
	\subfigure[ The band gap model at k-point $\frac{1}{2}(\Gamma \to X)$, for two different noise levels of  $\sigma=10^{-3}$ (left) and $\sigma=10^{-4}$ (right). \label{subfig:Mconv-gap}]{
		\includegraphics[]{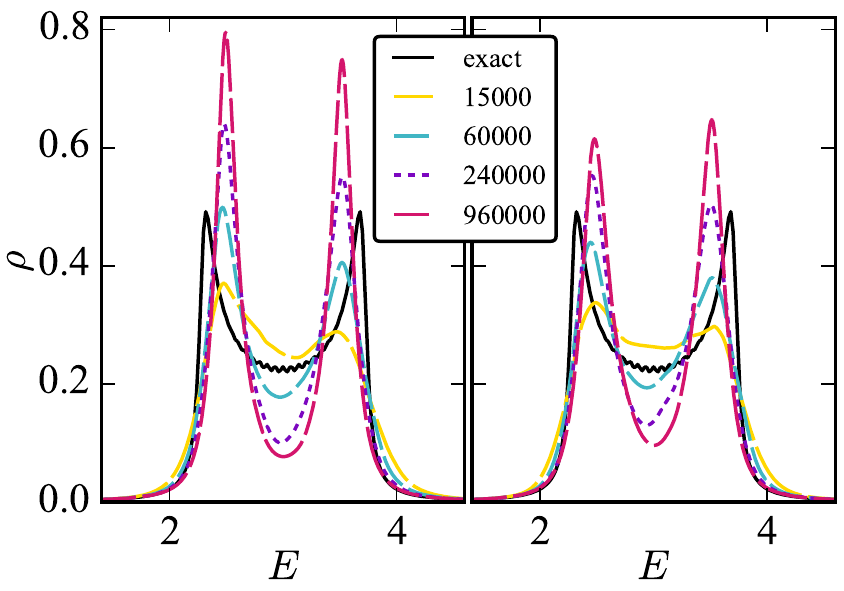}  
	}
	
	\vspace{0.7cm}
	
	\subfigure[ The doped Hubbard model at the $M$-point and Matsubara noise level $\sigma=10^{-4}$. \label{subfig:Mconv-doped}]{
		%\vspace{5.0cm}
		\includegraphics[]{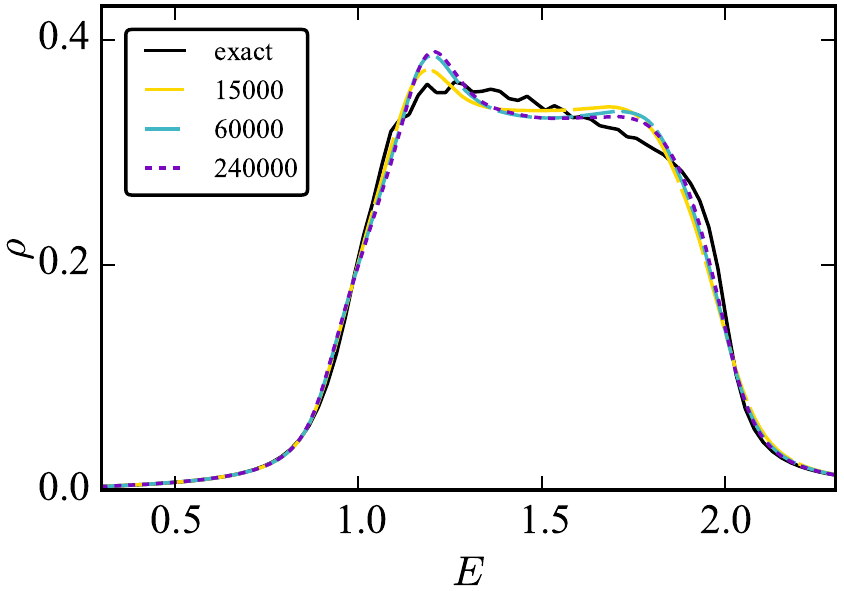}  
	}	
\caption{The dependence on the number of iterations (see legend) performed with the Mishchenko method.}
\label{fig:Mconv}
\end{figure}

\begin{figure}
	\includegraphics[]{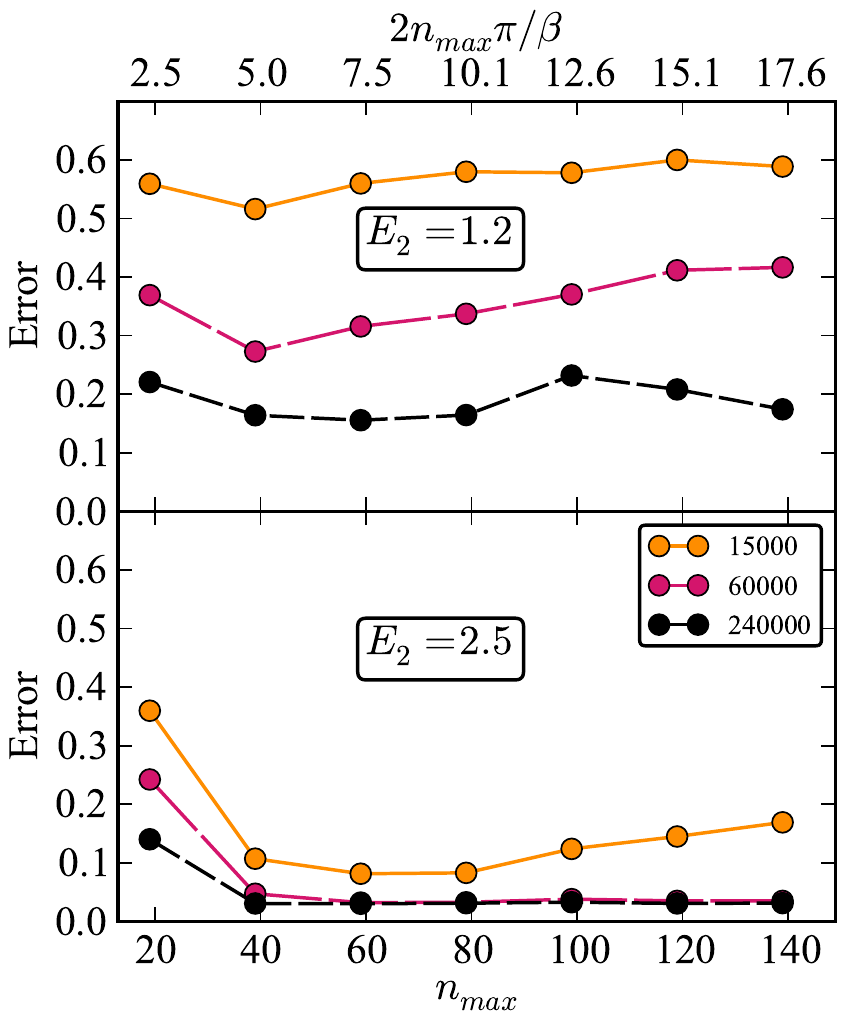}	
	\caption{Integrated real axis error for the two-pole model as a function of the highest Matsubara point index $n_\text{max}$ used in Mishchenko's method. Different number of local iterations are investigated and are indicated by the legend. Upper panel $a_1=0.1$, $a_2=0.335663$, $E_1=0.7$ and $E_2=1.2$, lower panel $a_1=0.1$, $a_2=0.7$, $E_1=0.7$ and $E_2=2.5$, as defined in Eq.~\ref{eq:two-pole}. Matsubara noise level $\sigma=10^{-4}$ is used.}
	\label{fig:Mishchenko_local}
\end{figure}

\section{Imposing mirror symmetry\label{sec:impose_mirrorsym}}
The MEM, the NNT and the NNLS method conserve the symmetry of the spectra $\rho(E)=-\rho(-E)$, since they use the symmetric definition of the Hilbert transform, Eq.\eqref{hilbert4bosons}.
The Pad\'e method does not have such symmetry imposed. For fermionic one-particle Greens functions and self-energies with zero imaginary part at the origin of the complex plane, it is shown~\cite{Schott15} that the continuation improves by taking a few negative Matsubara frequencies into account to impose the mirror symmetry. 
Such a procedure, with a single negative Matsubara frequency, has also been used for the optical conductivity~\cite{PhysRevB.82.165125}. We have tried the same scheme for the Pad\'e continuation of the bosonic functions in this paper, but in most cases imposing mirror symmetry actually gave a worse continuation.
In particular, when the exact spectrum has a broad, rather flat spectrum, such as at the $M$-point of the doped Hubbard model, imposing the symmetry can lead to spurious sharp features. Broad modes in the spectrum are related to branch cuts in the complex function $\chi(z)$. The Pad\'e approximant uses a finite number of poles to simulate $\chi(z)$. When the mirror symmetry is imposed on the Pad\'e approximant, the poles move closer to the real axis, which results in sharper spectral features. Hence broad features are harder to accurately simulate by imposing mirror symmetry.  

\section{Convergence of Mishchenko's sampling method\label{app:conv_mish}}
The stochastic sampling method by Mishchenko {\itshape{et al.}}~\cite{PhysRevB.62.6317} consists of many iterative steps. In every step, the configuration of the rectangles is altered. In the main text results with 15000 iterations are presented, which is a suitable setup to obtain reasonable results in a decent computational time (30 cpu hours using 128 cpus). However, in principle the method should be converged carefully with respect to the number of iterations. In Fig.~\ref{fig:Mconv} we increase that number by a factor of 4 several times, and show the resulting spectral function. For the two-pole model (see Fig.~\ref{fig:2pole:spectra}) it is clear that increasing the number of iterations improves the resulting spectrum, see Fig.~\ref{subfig:Mconv-2p}. However for a more complicated spectrum, such as the band gap model at the k-point $\frac{1}{2}(\Gamma \to X)$ (see Fig.~\ref{subfig:GX}), convergence with respect to the number of iterations does not lead to an improvement, see Fig.~\ref{subfig:Mconv-gap}. This issue is particularly pronounced when the input noise is big. The tendency to produce too sharp features resembles the overfitting occurring in the NNLS method. 
For the doped Hubbard model at the $M$-point (see left panel in Fig.~\ref{fig:RPA:noise:Pade_NNLS_NNT_MEM_2_inset}) the spectra converge already for 15000 iterations, see Fig.~\ref{subfig:Mconv-doped}.  
In summary, the optimal number of iterations is highly dependent on the function under investigation. With increasing number of iterations, the computational cost increases as well as the risk of overfitting to the input Matsubara data. Hence care should be taken when choosing the number of iterations.

Further, It is interesting to repeat the analysis on the accuracy of the continuation with respect to the number of Matsubara points (Fig.~\ref{fig:2pole:error}) while varying the number of local updates. These results are illustrated in Fig.~\ref{fig:Mishchenko_local}.  While increasing the number of local updates leads to a better continuation, the general trend reported in Fig.~\ref{fig:Mishchenko_local} is very similar to Fig.~\ref{fig:2pole:error}. Increasing the number of Matsubara frequencies the quality of the continuation reaches saturation very quickly and with a threshold that does not seem to depend on the number of local updates. A posteriori, this shows that the approach used in Fig.~\ref{fig:2pole:error}, i.e. keeping fixed the number of local updates while increasing the number of Matsubara points, is justifiable.

\newpage

%\bibliography{main}

\begin{thebibliography}{46}%
\makeatletter
\providecommand \@ifxundefined [1]{%
 \@ifx{#1\undefined}
}%
\providecommand \@ifnum [1]{%
 \ifnum #1\expandafter \@firstoftwo
 \else \expandafter \@secondoftwo
 \fi
}%
\providecommand \@ifx [1]{%
 \ifx #1\expandafter \@firstoftwo
 \else \expandafter \@secondoftwo
 \fi
}%
\providecommand \natexlab [1]{#1}%
\providecommand \enquote  [1]{``#1''}%
\providecommand \bibnamefont  [1]{#1}%
\providecommand \bibfnamefont [1]{#1}%
\providecommand \citenamefont [1]{#1}%
\providecommand \href@noop [0]{\@secondoftwo}%
\providecommand \href [0]{\begingroup \@sanitize@url \@href}%
\providecommand \@href[1]{\@@startlink{#1}\@@href}%
\providecommand \@@href[1]{\endgroup#1\@@endlink}%
\providecommand \@sanitize@url [0]{\catcode `\\12\catcode `\$12\catcode
  `\&12\catcode `\#12\catcode `\^12\catcode `\_12\catcode `\%12\relax}%
\providecommand \@@startlink[1]{}%
\providecommand \@@endlink[0]{}%
\providecommand \url  [0]{\begingroup\@sanitize@url \@url }%
\providecommand \@url [1]{\endgroup\@href {#1}{\urlprefix }}%
\providecommand \urlprefix  [0]{URL }%
\providecommand \Eprint [0]{\href }%
\providecommand \doibase [0]{http://dx.doi.org/}%
\providecommand \selectlanguage [0]{\@gobble}%
\providecommand \bibinfo  [0]{\@secondoftwo}%
\providecommand \bibfield  [0]{\@secondoftwo}%
\providecommand \translation [1]{[#1]}%
\providecommand \BibitemOpen [0]{}%
\providecommand \bibitemStop [0]{}%
\providecommand \bibitemNoStop [0]{.\EOS\space}%
\providecommand \EOS [0]{\spacefactor3000\relax}%
\providecommand \BibitemShut  [1]{\csname bibitem#1\endcsname}%
\let\auto@bib@innerbib\@empty
%</preamble>
\bibitem [{\citenamefont {Anisimov}\ and\ \citenamefont
  {Izyumov}(2010)}]{anisimov_book}%
  \BibitemOpen
  \bibfield  {author} {\bibinfo {author} {\bibfnamefont {V.}~\bibnamefont
  {Anisimov}}\ and\ \bibinfo {author} {\bibfnamefont {Y.}~\bibnamefont
  {Izyumov}},\ }\href@noop {} {\emph {\bibinfo {title} {Electronic Structure of
  Strongly Correlated Materials}}},\ edited by\ \bibinfo {editor}
  {\bibfnamefont {M.}~\bibnamefont {Cardona}},\ Springer Series in Solid-State
  Sciences\ (\bibinfo  {publisher} {Springer-Verlag},\ \bibinfo {address}
  {Berlin Heidelberg},\ \bibinfo {year} {2010})\BibitemShut {NoStop}%
\bibitem [{\citenamefont {Metzner}\ and\ \citenamefont
  {Vollhardt}(1989)}]{Metzner89}%
  \BibitemOpen
  \bibfield  {author} {\bibinfo {author} {\bibfnamefont {W.}~\bibnamefont
  {Metzner}}\ and\ \bibinfo {author} {\bibfnamefont {D.}~\bibnamefont
  {Vollhardt}},\ }\href {\doibase 10.1103/PhysRevLett.62.324} {\bibfield
  {journal} {\bibinfo  {journal} {Phys. Rev. Lett.}\ }\textbf {\bibinfo
  {volume} {62}},\ \bibinfo {pages} {324} (\bibinfo {year} {1989})}\BibitemShut
  {NoStop}%
\bibitem [{\citenamefont {Georges}\ \emph {et~al.}(1996)\citenamefont
  {Georges}, \citenamefont {Kotliar}, \citenamefont {Krauth},\ and\
  \citenamefont {Rozenberg}}]{Georges96}%
  \BibitemOpen
  \bibfield  {author} {\bibinfo {author} {\bibfnamefont {A.}~\bibnamefont
  {Georges}}, \bibinfo {author} {\bibfnamefont {G.}~\bibnamefont {Kotliar}},
  \bibinfo {author} {\bibfnamefont {W.}~\bibnamefont {Krauth}}, \ and\ \bibinfo
  {author} {\bibfnamefont {M.~J.}\ \bibnamefont {Rozenberg}},\ }\href {\doibase
  10.1103/RevModPhys.68.13} {\bibfield  {journal} {\bibinfo  {journal} {Rev.
  Mod. Phys.}\ }\textbf {\bibinfo {volume} {68}},\ \bibinfo {pages} {13}
  (\bibinfo {year} {1996})}\BibitemShut {NoStop}%
\bibitem [{\citenamefont {Lichtenstein}\ and\ \citenamefont
  {Katsnelson}(1998)}]{Lichtenstein98}%
  \BibitemOpen
  \bibfield  {author} {\bibinfo {author} {\bibfnamefont {A.~I.}\ \bibnamefont
  {Lichtenstein}}\ and\ \bibinfo {author} {\bibfnamefont {M.~I.}\ \bibnamefont
  {Katsnelson}},\ }\href {\doibase 10.1103/PhysRevB.57.6884} {\bibfield
  {journal} {\bibinfo  {journal} {Phys. Rev. B}\ }\textbf {\bibinfo {volume}
  {57}},\ \bibinfo {pages} {6884} (\bibinfo {year} {1998})}\BibitemShut
  {NoStop}%
\bibitem [{\citenamefont {Kotliar}\ \emph {et~al.}(2006)\citenamefont
  {Kotliar}, \citenamefont {Savrasov}, \citenamefont {Haule}, \citenamefont
  {Oudovenko}, \citenamefont {Parcollet},\ and\ \citenamefont
  {Marianetti}}]{Kotliar06}%
  \BibitemOpen
  \bibfield  {author} {\bibinfo {author} {\bibfnamefont {G.}~\bibnamefont
  {Kotliar}}, \bibinfo {author} {\bibfnamefont {S.~Y.}\ \bibnamefont
  {Savrasov}}, \bibinfo {author} {\bibfnamefont {K.}~\bibnamefont {Haule}},
  \bibinfo {author} {\bibfnamefont {V.~S.}\ \bibnamefont {Oudovenko}}, \bibinfo
  {author} {\bibfnamefont {O.}~\bibnamefont {Parcollet}}, \ and\ \bibinfo
  {author} {\bibfnamefont {C.~A.}\ \bibnamefont {Marianetti}},\ }\href
  {\doibase 10.1103/RevModPhys.78.865} {\bibfield  {journal} {\bibinfo
  {journal} {Rev. Mod. Phys.}\ }\textbf {\bibinfo {volume} {78}},\ \bibinfo
  {pages} {865} (\bibinfo {year} {2006})}\BibitemShut {NoStop}%
\bibitem [{\citenamefont {Negele}\ and\ \citenamefont
  {Orland}(1988)}]{NegeleOrland}%
  \BibitemOpen
  \bibfield  {author} {\bibinfo {author} {\bibfnamefont {J.~W.}\ \bibnamefont
  {Negele}}\ and\ \bibinfo {author} {\bibfnamefont {H.}~\bibnamefont
  {Orland}},\ }\href@noop {} {\emph {\bibinfo {title} {Quantum Many-Particle
  Systems}}}\ (\bibinfo  {publisher} {Addison-Wesley Publishing Company},\
  \bibinfo {year} {1988})\BibitemShut {NoStop}%
\bibitem [{\citenamefont {Mahan}(2000)}]{Mahan}%
  \BibitemOpen
  \bibfield  {author} {\bibinfo {author} {\bibfnamefont {G.~D.}\ \bibnamefont
  {Mahan}},\ }\href@noop {} {\emph {\bibinfo {title} {Many Particle Physics,
  Third Edition}}}\ (\bibinfo  {publisher} {Plenum},\ \bibinfo {address} {New
  York},\ \bibinfo {year} {2000})\BibitemShut {NoStop}%
\bibitem [{\citenamefont {Boehnke}\ and\ \citenamefont
  {Lechermann}(2012)}]{Boehnke12}%
  \BibitemOpen
  \bibfield  {author} {\bibinfo {author} {\bibfnamefont {L.}~\bibnamefont
  {Boehnke}}\ and\ \bibinfo {author} {\bibfnamefont {F.}~\bibnamefont
  {Lechermann}},\ }\href {\doibase 10.1103/PhysRevB.85.115128} {\bibfield
  {journal} {\bibinfo  {journal} {Phys. Rev. B}\ }\textbf {\bibinfo {volume}
  {85}},\ \bibinfo {pages} {115128} (\bibinfo {year} {2012})}\BibitemShut
  {NoStop}%
\bibitem [{\citenamefont {Ayral}\ \emph {et~al.}(2012)\citenamefont {Ayral},
  \citenamefont {Werner},\ and\ \citenamefont {Biermann}}]{Ayral12}%
  \BibitemOpen
  \bibfield  {author} {\bibinfo {author} {\bibfnamefont {T.}~\bibnamefont
  {Ayral}}, \bibinfo {author} {\bibfnamefont {P.}~\bibnamefont {Werner}}, \
  and\ \bibinfo {author} {\bibfnamefont {S.}~\bibnamefont {Biermann}},\ }\href
  {\doibase 10.1103/PhysRevLett.109.226401} {\bibfield  {journal} {\bibinfo
  {journal} {Phys. Rev. Lett.}\ }\textbf {\bibinfo {volume} {109}},\ \bibinfo
  {pages} {226401} (\bibinfo {year} {2012})}\BibitemShut {NoStop}%
\bibitem [{\citenamefont {Hansmann}\ \emph {et~al.}(2013)\citenamefont
  {Hansmann}, \citenamefont {Ayral}, \citenamefont {Vaugier}, \citenamefont
  {Werner},\ and\ \citenamefont {Biermann}}]{Hansmann13}%
  \BibitemOpen
  \bibfield  {author} {\bibinfo {author} {\bibfnamefont {P.}~\bibnamefont
  {Hansmann}}, \bibinfo {author} {\bibfnamefont {T.}~\bibnamefont {Ayral}},
  \bibinfo {author} {\bibfnamefont {L.}~\bibnamefont {Vaugier}}, \bibinfo
  {author} {\bibfnamefont {P.}~\bibnamefont {Werner}}, \ and\ \bibinfo {author}
  {\bibfnamefont {S.}~\bibnamefont {Biermann}},\ }\href {\doibase
  10.1103/PhysRevLett.110.166401} {\bibfield  {journal} {\bibinfo  {journal}
  {Phys. Rev. Lett.}\ }\textbf {\bibinfo {volume} {110}},\ \bibinfo {pages}
  {166401} (\bibinfo {year} {2013})}\BibitemShut {NoStop}%
\bibitem [{\citenamefont {Ayral}\ \emph {et~al.}(2013)\citenamefont {Ayral},
  \citenamefont {Biermann},\ and\ \citenamefont {Werner}}]{Ayral13}%
  \BibitemOpen
  \bibfield  {author} {\bibinfo {author} {\bibfnamefont {T.}~\bibnamefont
  {Ayral}}, \bibinfo {author} {\bibfnamefont {S.}~\bibnamefont {Biermann}}, \
  and\ \bibinfo {author} {\bibfnamefont {P.}~\bibnamefont {Werner}},\ }\href
  {\doibase 10.1103/PhysRevB.87.125149} {\bibfield  {journal} {\bibinfo
  {journal} {Phys. Rev. B}\ }\textbf {\bibinfo {volume} {87}},\ \bibinfo
  {pages} {125149} (\bibinfo {year} {2013})}\BibitemShut {NoStop}%
\bibitem [{\citenamefont {Huang}\ \emph {et~al.}(2014)\citenamefont {Huang},
  \citenamefont {Ayral}, \citenamefont {Biermann},\ and\ \citenamefont
  {Werner}}]{Huang14}%
  \BibitemOpen
  \bibfield  {author} {\bibinfo {author} {\bibfnamefont {L.}~\bibnamefont
  {Huang}}, \bibinfo {author} {\bibfnamefont {T.}~\bibnamefont {Ayral}},
  \bibinfo {author} {\bibfnamefont {S.}~\bibnamefont {Biermann}}, \ and\
  \bibinfo {author} {\bibfnamefont {P.}~\bibnamefont {Werner}},\ }\href
  {\doibase 10.1103/PhysRevB.90.195114} {\bibfield  {journal} {\bibinfo
  {journal} {Phys. Rev. B}\ }\textbf {\bibinfo {volume} {90}},\ \bibinfo
  {pages} {195114} (\bibinfo {year} {2014})}\BibitemShut {NoStop}%
\bibitem [{\citenamefont {van Loon}\ \emph
  {et~al.}(2014{\natexlab{a}})\citenamefont {van Loon}, \citenamefont
  {Hafermann}, \citenamefont {Lichtenstein}, \citenamefont {Rubtsov},\ and\
  \citenamefont {Katsnelson}}]{vanLoon14}%
  \BibitemOpen
  \bibfield  {author} {\bibinfo {author} {\bibfnamefont {E.~G. C.~P.}\
  \bibnamefont {van Loon}}, \bibinfo {author} {\bibfnamefont {H.}~\bibnamefont
  {Hafermann}}, \bibinfo {author} {\bibfnamefont {A.~I.}\ \bibnamefont
  {Lichtenstein}}, \bibinfo {author} {\bibfnamefont {A.~N.}\ \bibnamefont
  {Rubtsov}}, \ and\ \bibinfo {author} {\bibfnamefont {M.~I.}\ \bibnamefont
  {Katsnelson}},\ }\href {\doibase 10.1103/PhysRevLett.113.246407} {\bibfield
  {journal} {\bibinfo  {journal} {Phys. Rev. Lett.}\ }\textbf {\bibinfo
  {volume} {113}},\ \bibinfo {pages} {246407} (\bibinfo {year}
  {2014}{\natexlab{a}})}\BibitemShut {NoStop}%
\bibitem [{\citenamefont {Hafermann}\ \emph {et~al.}(2014)\citenamefont
  {Hafermann}, \citenamefont {van Loon}, \citenamefont {Katsnelson},
  \citenamefont {Lichtenstein},\ and\ \citenamefont
  {Parcollet}}]{Hafermann14-2}%
  \BibitemOpen
  \bibfield  {author} {\bibinfo {author} {\bibfnamefont {H.}~\bibnamefont
  {Hafermann}}, \bibinfo {author} {\bibfnamefont {E.~G. C.~P.}\ \bibnamefont
  {van Loon}}, \bibinfo {author} {\bibfnamefont {M.~I.}\ \bibnamefont
  {Katsnelson}}, \bibinfo {author} {\bibfnamefont {A.~I.}\ \bibnamefont
  {Lichtenstein}}, \ and\ \bibinfo {author} {\bibfnamefont {O.}~\bibnamefont
  {Parcollet}},\ }\href {\doibase 10.1103/PhysRevB.90.235105} {\bibfield
  {journal} {\bibinfo  {journal} {Phys. Rev. B}\ }\textbf {\bibinfo {volume}
  {90}},\ \bibinfo {pages} {235105} (\bibinfo {year} {2014})}\BibitemShut
  {NoStop}%
\bibitem [{\citenamefont {Galler}\ \emph {et~al.}(2015)\citenamefont {Galler},
  \citenamefont {Taranto}, \citenamefont {Wallerberger}, \citenamefont
  {Kaltak}, \citenamefont {Kresse}, \citenamefont {Sangiovanni}, \citenamefont
  {Toschi},\ and\ \citenamefont {Held}}]{Galler15}%
  \BibitemOpen
  \bibfield  {author} {\bibinfo {author} {\bibfnamefont {A.}~\bibnamefont
  {Galler}}, \bibinfo {author} {\bibfnamefont {C.}~\bibnamefont {Taranto}},
  \bibinfo {author} {\bibfnamefont {M.}~\bibnamefont {Wallerberger}}, \bibinfo
  {author} {\bibfnamefont {M.}~\bibnamefont {Kaltak}}, \bibinfo {author}
  {\bibfnamefont {G.}~\bibnamefont {Kresse}}, \bibinfo {author} {\bibfnamefont
  {G.}~\bibnamefont {Sangiovanni}}, \bibinfo {author} {\bibfnamefont
  {A.}~\bibnamefont {Toschi}}, \ and\ \bibinfo {author} {\bibfnamefont
  {K.}~\bibnamefont {Held}},\ }\href {\doibase 10.1103/PhysRevB.92.205132}
  {\bibfield  {journal} {\bibinfo  {journal} {Phys. Rev. B}\ }\textbf {\bibinfo
  {volume} {92}},\ \bibinfo {pages} {205132} (\bibinfo {year}
  {2015})}\BibitemShut {NoStop}%
\bibitem [{\citenamefont {Mishchenko}\ \emph {et~al.}(2015)\citenamefont
  {Mishchenko}, \citenamefont {Nagaosa}, \citenamefont {De~Filippis},
  \citenamefont {de~Candia},\ and\ \citenamefont
  {Cataudella}}]{PhysRevLett.114.146401}%
  \BibitemOpen
  \bibfield  {author} {\bibinfo {author} {\bibfnamefont {A.~S.}\ \bibnamefont
  {Mishchenko}}, \bibinfo {author} {\bibfnamefont {N.}~\bibnamefont {Nagaosa}},
  \bibinfo {author} {\bibfnamefont {G.}~\bibnamefont {De~Filippis}}, \bibinfo
  {author} {\bibfnamefont {A.}~\bibnamefont {de~Candia}}, \ and\ \bibinfo
  {author} {\bibfnamefont {V.}~\bibnamefont {Cataudella}},\ }\href {\doibase
  10.1103/PhysRevLett.114.146401} {\bibfield  {journal} {\bibinfo  {journal}
  {Phys. Rev. Lett.}\ }\textbf {\bibinfo {volume} {114}},\ \bibinfo {pages}
  {146401} (\bibinfo {year} {2015})}\BibitemShut {NoStop}%
\bibitem [{\citenamefont {Katanin}\ \emph {et~al.}(2010)\citenamefont
  {Katanin}, \citenamefont {Poteryaev}, \citenamefont {Efremov}, \citenamefont
  {Shorikov}, \citenamefont {Skornyakov}, \citenamefont {Korotin},\ and\
  \citenamefont {Anisimov}}]{katanin10}%
  \BibitemOpen
  \bibfield  {author} {\bibinfo {author} {\bibfnamefont {A.~A.}\ \bibnamefont
  {Katanin}}, \bibinfo {author} {\bibfnamefont {A.~I.}\ \bibnamefont
  {Poteryaev}}, \bibinfo {author} {\bibfnamefont {A.~V.}\ \bibnamefont
  {Efremov}}, \bibinfo {author} {\bibfnamefont {A.~O.}\ \bibnamefont
  {Shorikov}}, \bibinfo {author} {\bibfnamefont {S.~L.}\ \bibnamefont
  {Skornyakov}}, \bibinfo {author} {\bibfnamefont {M.~A.}\ \bibnamefont
  {Korotin}}, \ and\ \bibinfo {author} {\bibfnamefont {V.~I.}\ \bibnamefont
  {Anisimov}},\ }\href {\doibase 10.1103/PhysRevB.81.045117} {\bibfield
  {journal} {\bibinfo  {journal} {Phys. Rev. B}\ }\textbf {\bibinfo {volume}
  {81}},\ \bibinfo {pages} {045117} (\bibinfo {year} {2010})}\BibitemShut
  {NoStop}%
\bibitem [{\citenamefont {Boehnke}\ \emph {et~al.}(2011)\citenamefont
  {Boehnke}, \citenamefont {Hafermann}, \citenamefont {Ferrero}, \citenamefont
  {Lechermann},\ and\ \citenamefont {Parcollet}}]{Boehnke11}%
  \BibitemOpen
  \bibfield  {author} {\bibinfo {author} {\bibfnamefont {L.}~\bibnamefont
  {Boehnke}}, \bibinfo {author} {\bibfnamefont {H.}~\bibnamefont {Hafermann}},
  \bibinfo {author} {\bibfnamefont {M.}~\bibnamefont {Ferrero}}, \bibinfo
  {author} {\bibfnamefont {F.}~\bibnamefont {Lechermann}}, \ and\ \bibinfo
  {author} {\bibfnamefont {O.}~\bibnamefont {Parcollet}},\ }\href {\doibase
  10.1103/PhysRevB.84.075145} {\bibfield  {journal} {\bibinfo  {journal} {Phys.
  Rev. B}\ }\textbf {\bibinfo {volume} {84}},\ \bibinfo {pages} {075145}
  (\bibinfo {year} {2011})}\BibitemShut {NoStop}%
\bibitem [{\citenamefont {Si}\ and\ \citenamefont {Smith}(1996)}]{Si96}%
  \BibitemOpen
  \bibfield  {author} {\bibinfo {author} {\bibfnamefont {Q.}~\bibnamefont
  {Si}}\ and\ \bibinfo {author} {\bibfnamefont {J.~L.}\ \bibnamefont {Smith}},\
  }\href {\doibase 10.1103/PhysRevLett.77.3391} {\bibfield  {journal} {\bibinfo
   {journal} {Phys. Rev. Lett.}\ }\textbf {\bibinfo {volume} {77}},\ \bibinfo
  {pages} {3391} (\bibinfo {year} {1996})}\BibitemShut {NoStop}%
\bibitem [{\citenamefont {Kajueter}(1996)}]{Kajueter96}%
  \BibitemOpen
  \bibfield  {author} {\bibinfo {author} {\bibfnamefont {H.}~\bibnamefont
  {Kajueter}},\ }\href@noop {} {Ph.D. thesis},\ \bibinfo  {school} {Rutgers
  University} (\bibinfo {year} {1996})\BibitemShut {NoStop}%
\bibitem [{\citenamefont {Smith}\ and\ \citenamefont {Si}(2000)}]{Smith00}%
  \BibitemOpen
  \bibfield  {author} {\bibinfo {author} {\bibfnamefont {J.~L.}\ \bibnamefont
  {Smith}}\ and\ \bibinfo {author} {\bibfnamefont {Q.}~\bibnamefont {Si}},\
  }\href {\doibase 10.1103/PhysRevB.61.5184} {\bibfield  {journal} {\bibinfo
  {journal} {Phys. Rev. B}\ }\textbf {\bibinfo {volume} {61}},\ \bibinfo
  {pages} {5184} (\bibinfo {year} {2000})}\BibitemShut {NoStop}%
\bibitem [{\citenamefont {Chitra}\ and\ \citenamefont
  {Kotliar}(2000)}]{Chitra00}%
  \BibitemOpen
  \bibfield  {author} {\bibinfo {author} {\bibfnamefont {R.}~\bibnamefont
  {Chitra}}\ and\ \bibinfo {author} {\bibfnamefont {G.}~\bibnamefont
  {Kotliar}},\ }\href {\doibase 10.1103/PhysRevLett.84.3678} {\bibfield
  {journal} {\bibinfo  {journal} {Phys. Rev. Lett.}\ }\textbf {\bibinfo
  {volume} {84}},\ \bibinfo {pages} {3678} (\bibinfo {year}
  {2000})}\BibitemShut {NoStop}%
\bibitem [{\citenamefont {Chitra}\ and\ \citenamefont
  {Kotliar}(2001)}]{Chitra01}%
  \BibitemOpen
  \bibfield  {author} {\bibinfo {author} {\bibfnamefont {R.}~\bibnamefont
  {Chitra}}\ and\ \bibinfo {author} {\bibfnamefont {G.}~\bibnamefont
  {Kotliar}},\ }\href {\doibase 10.1103/PhysRevB.63.115110} {\bibfield
  {journal} {\bibinfo  {journal} {Phys. Rev. B}\ }\textbf {\bibinfo {volume}
  {63}},\ \bibinfo {pages} {115110} (\bibinfo {year} {2001})}\BibitemShut
  {NoStop}%
\bibitem [{\citenamefont {Rubtsov}\ \emph {et~al.}(2012)\citenamefont
  {Rubtsov}, \citenamefont {Katsnelson},\ and\ \citenamefont
  {Lichtenstein}}]{Rubtsov12}%
  \BibitemOpen
  \bibfield  {author} {\bibinfo {author} {\bibfnamefont {A.~N.}\ \bibnamefont
  {Rubtsov}}, \bibinfo {author} {\bibfnamefont {M.~I.}\ \bibnamefont
  {Katsnelson}}, \ and\ \bibinfo {author} {\bibfnamefont {A.~I.}\ \bibnamefont
  {Lichtenstein}},\ }\href {\doibase 10.1016/j.aop.2012.01.002} {\bibfield
  {journal} {\bibinfo  {journal} {Annals of Physics}\ }\textbf {\bibinfo
  {volume} {327}},\ \bibinfo {pages} {1320} (\bibinfo {year}
  {2012})}\BibitemShut {NoStop}%
\bibitem [{\citenamefont {van Loon}\ \emph
  {et~al.}(2014{\natexlab{b}})\citenamefont {van Loon}, \citenamefont
  {Lichtenstein}, \citenamefont {Katsnelson}, \citenamefont {Parcollet},\ and\
  \citenamefont {Hafermann}}]{vanLoon14-2}%
  \BibitemOpen
  \bibfield  {author} {\bibinfo {author} {\bibfnamefont {E.~G. C.~P.}\
  \bibnamefont {van Loon}}, \bibinfo {author} {\bibfnamefont {A.~I.}\
  \bibnamefont {Lichtenstein}}, \bibinfo {author} {\bibfnamefont {M.~I.}\
  \bibnamefont {Katsnelson}}, \bibinfo {author} {\bibfnamefont
  {O.}~\bibnamefont {Parcollet}}, \ and\ \bibinfo {author} {\bibfnamefont
  {H.}~\bibnamefont {Hafermann}},\ }\href {\doibase 10.1103/PhysRevB.90.235135}
  {\bibfield  {journal} {\bibinfo  {journal} {Phys. Rev. B}\ }\textbf {\bibinfo
  {volume} {90}},\ \bibinfo {pages} {235135} (\bibinfo {year}
  {2014}{\natexlab{b}})}\BibitemShut {NoStop}%
\bibitem [{\citenamefont {van Loon}\ \emph {et~al.}(2015)\citenamefont {van
  Loon}, \citenamefont {Katsnelson},\ and\ \citenamefont
  {Lemeshko}}]{vanLoon15-2}%
  \BibitemOpen
  \bibfield  {author} {\bibinfo {author} {\bibfnamefont {E.~G. C.~P.}\
  \bibnamefont {van Loon}}, \bibinfo {author} {\bibfnamefont {M.~I.}\
  \bibnamefont {Katsnelson}}, \ and\ \bibinfo {author} {\bibfnamefont
  {M.}~\bibnamefont {Lemeshko}},\ }\href {\doibase 10.1103/PhysRevB.92.081106}
  {\bibfield  {journal} {\bibinfo  {journal} {Phys. Rev. B}\ }\textbf {\bibinfo
  {volume} {92}},\ \bibinfo {pages} {081106} (\bibinfo {year}
  {2015})}\BibitemShut {NoStop}%
\bibitem [{\citenamefont {Gunnarsson}\ \emph
  {et~al.}(2010{\natexlab{a}})\citenamefont {Gunnarsson}, \citenamefont
  {Haverkort},\ and\ \citenamefont {Sangiovanni}}]{PhysRevB.82.165125}%
  \BibitemOpen
  \bibfield  {author} {\bibinfo {author} {\bibfnamefont {O.}~\bibnamefont
  {Gunnarsson}}, \bibinfo {author} {\bibfnamefont {M.~W.}\ \bibnamefont
  {Haverkort}}, \ and\ \bibinfo {author} {\bibfnamefont {G.}~\bibnamefont
  {Sangiovanni}},\ }\href {\doibase 10.1103/PhysRevB.82.165125} {\bibfield
  {journal} {\bibinfo  {journal} {Phys. Rev. B}\ }\textbf {\bibinfo {volume}
  {82}},\ \bibinfo {pages} {165125} (\bibinfo {year}
  {2010}{\natexlab{a}})}\BibitemShut {NoStop}%
\bibitem [{\citenamefont {Gunnarsson}\ \emph
  {et~al.}(2010{\natexlab{b}})\citenamefont {Gunnarsson}, \citenamefont
  {Haverkort},\ and\ \citenamefont {Sangiovanni}}]{PhysRevB.81.155107}%
  \BibitemOpen
  \bibfield  {author} {\bibinfo {author} {\bibfnamefont {O.}~\bibnamefont
  {Gunnarsson}}, \bibinfo {author} {\bibfnamefont {M.~W.}\ \bibnamefont
  {Haverkort}}, \ and\ \bibinfo {author} {\bibfnamefont {G.}~\bibnamefont
  {Sangiovanni}},\ }\href {\doibase 10.1103/PhysRevB.81.155107} {\bibfield
  {journal} {\bibinfo  {journal} {Phys. Rev. B}\ }\textbf {\bibinfo {volume}
  {81}},\ \bibinfo {pages} {155107} (\bibinfo {year}
  {2010}{\natexlab{b}})}\BibitemShut {NoStop}%
\bibitem [{\citenamefont {Casula}\ \emph {et~al.}(2012)\citenamefont {Casula},
  \citenamefont {Rubtsov},\ and\ \citenamefont {Biermann}}]{Casula12}%
  \BibitemOpen
  \bibfield  {author} {\bibinfo {author} {\bibfnamefont {M.}~\bibnamefont
  {Casula}}, \bibinfo {author} {\bibfnamefont {A.}~\bibnamefont {Rubtsov}}, \
  and\ \bibinfo {author} {\bibfnamefont {S.}~\bibnamefont {Biermann}},\ }\href
  {\doibase 10.1103/PhysRevB.85.035115} {\bibfield  {journal} {\bibinfo
  {journal} {Phys. Rev. B}\ }\textbf {\bibinfo {volume} {85}},\ \bibinfo
  {pages} {035115} (\bibinfo {year} {2012})}\BibitemShut {NoStop}%
\bibitem [{\citenamefont {Silver}\ \emph {et~al.}(1990)\citenamefont {Silver},
  \citenamefont {Sivia},\ and\ \citenamefont {Gubernatis}}]{PhysRevB.41.2380}%
  \BibitemOpen
  \bibfield  {author} {\bibinfo {author} {\bibfnamefont {R.~N.}\ \bibnamefont
  {Silver}}, \bibinfo {author} {\bibfnamefont {D.~S.}\ \bibnamefont {Sivia}}, \
  and\ \bibinfo {author} {\bibfnamefont {J.~E.}\ \bibnamefont {Gubernatis}},\
  }\href {\doibase 10.1103/PhysRevB.41.2380} {\bibfield  {journal} {\bibinfo
  {journal} {Phys. Rev. B}\ }\textbf {\bibinfo {volume} {41}},\ \bibinfo
  {pages} {2380} (\bibinfo {year} {1990})}\BibitemShut {NoStop}%
\bibitem [{\citenamefont {Jarrell}\ and\ \citenamefont
  {Gubernatis}(1996)}]{Jarrell1996133}%
  \BibitemOpen
  \bibfield  {author} {\bibinfo {author} {\bibfnamefont {M.}~\bibnamefont
  {Jarrell}}\ and\ \bibinfo {author} {\bibfnamefont {J.}~\bibnamefont
  {Gubernatis}},\ }\href {\doibase
  http://dx.doi.org/10.1016/0370-1573(95)00074-7} {\bibfield  {journal}
  {\bibinfo  {journal} {Physics Reports}\ }\textbf {\bibinfo {volume} {269}},\
  \bibinfo {pages} {133 } (\bibinfo {year} {1996})}\BibitemShut {NoStop}%
\bibitem [{\citenamefont {Bryan}(1990)}]{bryans}%
  \BibitemOpen
  \bibfield  {author} {\bibinfo {author} {\bibfnamefont {R.}~\bibnamefont
  {Bryan}},\ }\href {\doibase 10.1007/BF02427376} {\bibfield  {journal}
  {\bibinfo  {journal} {European Biophysics Journal}\ }\textbf {\bibinfo
  {volume} {18}},\ \bibinfo {pages} {165} (\bibinfo {year} {1990})}\BibitemShut
  {NoStop}%
\bibitem [{\citenamefont {Gubernatis}\ \emph {et~al.}(1991)\citenamefont
  {Gubernatis}, \citenamefont {Jarrell}, \citenamefont {Silver},\ and\
  \citenamefont {Sivia}}]{PhysRevB.44.6011}%
  \BibitemOpen
  \bibfield  {author} {\bibinfo {author} {\bibfnamefont {J.~E.}\ \bibnamefont
  {Gubernatis}}, \bibinfo {author} {\bibfnamefont {M.}~\bibnamefont {Jarrell}},
  \bibinfo {author} {\bibfnamefont {R.~N.}\ \bibnamefont {Silver}}, \ and\
  \bibinfo {author} {\bibfnamefont {D.~S.}\ \bibnamefont {Sivia}},\ }\href
  {\doibase 10.1103/PhysRevB.44.6011} {\bibfield  {journal} {\bibinfo
  {journal} {Phys. Rev. B}\ }\textbf {\bibinfo {volume} {44}},\ \bibinfo
  {pages} {6011} (\bibinfo {year} {1991})}\BibitemShut {NoStop}%
\bibitem [{\citenamefont {Fuchs}\ \emph {et~al.}(2010)\citenamefont {Fuchs},
  \citenamefont {Pruschke},\ and\ \citenamefont
  {Jarrell}}]{PhysRevE.81.056701}%
  \BibitemOpen
  \bibfield  {author} {\bibinfo {author} {\bibfnamefont {S.}~\bibnamefont
  {Fuchs}}, \bibinfo {author} {\bibfnamefont {T.}~\bibnamefont {Pruschke}}, \
  and\ \bibinfo {author} {\bibfnamefont {M.}~\bibnamefont {Jarrell}},\ }\href
  {\doibase 10.1103/PhysRevE.81.056701} {\bibfield  {journal} {\bibinfo
  {journal} {Phys. Rev. E}\ }\textbf {\bibinfo {volume} {81}},\ \bibinfo
  {pages} {056701} (\bibinfo {year} {2010})}\BibitemShut {NoStop}%
\bibitem [{\citenamefont {Dirks}\ \emph {et~al.}(2010)\citenamefont {Dirks},
  \citenamefont {Werner}, \citenamefont {Jarrell},\ and\ \citenamefont
  {Pruschke}}]{PhysRevE.82.026701}%
  \BibitemOpen
  \bibfield  {author} {\bibinfo {author} {\bibfnamefont {A.}~\bibnamefont
  {Dirks}}, \bibinfo {author} {\bibfnamefont {P.}~\bibnamefont {Werner}},
  \bibinfo {author} {\bibfnamefont {M.}~\bibnamefont {Jarrell}}, \ and\
  \bibinfo {author} {\bibfnamefont {T.}~\bibnamefont {Pruschke}},\ }\href
  {\doibase 10.1103/PhysRevE.82.026701} {\bibfield  {journal} {\bibinfo
  {journal} {Phys. Rev. E}\ }\textbf {\bibinfo {volume} {82}},\ \bibinfo
  {pages} {026701} (\bibinfo {year} {2010})}\BibitemShut {NoStop}%
\bibitem [{\citenamefont {Lawson}\ and\ \citenamefont {Hanson}(1995)}]{NNLS}%
  \BibitemOpen
  \bibfield  {author} {\bibinfo {author} {\bibfnamefont {C.~L.}\ \bibnamefont
  {Lawson}}\ and\ \bibinfo {author} {\bibfnamefont {R.~J.}\ \bibnamefont
  {Hanson}},\ }\href@noop {} {\emph {\bibinfo {title} {Solving Least Squares
  Problems}}}\ (\bibinfo  {publisher} {Society for Industrial and Applied
  Mathematics Philadelphia},\ \bibinfo {address} {Philadelphia},\ \bibinfo
  {year} {1995})\BibitemShut {NoStop}%
\bibitem [{\citenamefont {Hansen}(2000)}]{L-curve}%
  \BibitemOpen
  \bibfield  {author} {\bibinfo {author} {\bibfnamefont {P.~C.}\ \bibnamefont
  {Hansen}},\ }in\ \href@noop {} {\emph {\bibinfo {booktitle} {Computational
  Inverse Problems in Electrocardiology, ed. P. Johnston, Advances in
  Computational Bioengineering}}}\ (\bibinfo  {publisher} {WIT Press},\
  \bibinfo {year} {2000})\ pp.\ \bibinfo {pages} {119--142}\BibitemShut
  {NoStop}%
\bibitem [{\citenamefont {Beach}\ \emph {et~al.}(2000)\citenamefont {Beach},
  \citenamefont {Gooding},\ and\ \citenamefont {Marsiglio}}]{PhysRevB.61.5147}%
  \BibitemOpen
  \bibfield  {author} {\bibinfo {author} {\bibfnamefont {K.~S.~D.}\
  \bibnamefont {Beach}}, \bibinfo {author} {\bibfnamefont {R.~J.}\ \bibnamefont
  {Gooding}}, \ and\ \bibinfo {author} {\bibfnamefont {F.}~\bibnamefont
  {Marsiglio}},\ }\href {\doibase 10.1103/PhysRevB.61.5147} {\bibfield
  {journal} {\bibinfo  {journal} {Phys. Rev. B}\ }\textbf {\bibinfo {volume}
  {61}},\ \bibinfo {pages} {5147} (\bibinfo {year} {2000})}\BibitemShut
  {NoStop}%
\bibitem [{\citenamefont {Vidberg}\ and\ \citenamefont
  {Serene}(1977)}]{Vidberg77}%
  \BibitemOpen
  \bibfield  {author} {\bibinfo {author} {\bibfnamefont {H.~J.}\ \bibnamefont
  {Vidberg}}\ and\ \bibinfo {author} {\bibfnamefont {J.~W.}\ \bibnamefont
  {Serene}},\ }\href {\doibase 10.1007/BF00655090} {\bibfield  {journal}
  {\bibinfo  {journal} {J. Low Temp. Phys.}\ }\textbf {\bibinfo {volume}
  {29}},\ \bibinfo {pages} {179} (\bibinfo {year} {1977})}\BibitemShut
  {NoStop}%
\bibitem [{\citenamefont {Mishchenko}\ \emph {et~al.}(2000)\citenamefont
  {Mishchenko}, \citenamefont {Prokof'ev}, \citenamefont {Sakamoto},\ and\
  \citenamefont {Svistunov}}]{PhysRevB.62.6317}%
  \BibitemOpen
  \bibfield  {author} {\bibinfo {author} {\bibfnamefont {A.~S.}\ \bibnamefont
  {Mishchenko}}, \bibinfo {author} {\bibfnamefont {N.~V.}\ \bibnamefont
  {Prokof'ev}}, \bibinfo {author} {\bibfnamefont {A.}~\bibnamefont {Sakamoto}},
  \ and\ \bibinfo {author} {\bibfnamefont {B.~V.}\ \bibnamefont {Svistunov}},\
  }\href {\doibase 10.1103/PhysRevB.62.6317} {\bibfield  {journal} {\bibinfo
  {journal} {Phys. Rev. B}\ }\textbf {\bibinfo {volume} {62}},\ \bibinfo
  {pages} {6317} (\bibinfo {year} {2000})}\BibitemShut {NoStop}%
\bibitem [{\citenamefont {Mishchenko}(2012)}]{MishchenkoJulich}%
  \BibitemOpen
  \bibfield  {author} {\bibinfo {author} {\bibfnamefont {A.~S.}\ \bibnamefont
  {Mishchenko}},\ }in\ \href {http://juser.fz-juelich.de/record/22915} {\emph
  {\bibinfo {booktitle} {Verlag des Forschungszentrum Jülich, 2012. -
  978-3-89336-796-2}}},\ \bibinfo {editor} {edited by\ \bibinfo {editor}
  {\bibfnamefont {E.}~\bibnamefont {Pavarini}}, \bibinfo {editor}
  {\bibfnamefont {E.}~\bibnamefont {Koch}}, \bibinfo {editor} {\bibfnamefont
  {F.}~\bibnamefont {Anders}}, \ and\ \bibinfo {editor} {\bibfnamefont
  {M.}~\bibnamefont {Jarrell}}}\ (\bibinfo {year} {2012})\BibitemShut {NoStop}%
\bibitem [{\citenamefont {Gull}\ \emph {et~al.}(2011)\citenamefont {Gull},
  \citenamefont {Millis}, \citenamefont {Lichtenstein}, \citenamefont
  {Rubtsov}, \citenamefont {Troyer},\ and\ \citenamefont {Werner}}]{Gull11}%
  \BibitemOpen
  \bibfield  {author} {\bibinfo {author} {\bibfnamefont {E.}~\bibnamefont
  {Gull}}, \bibinfo {author} {\bibfnamefont {A.~J.}\ \bibnamefont {Millis}},
  \bibinfo {author} {\bibfnamefont {A.~I.}\ \bibnamefont {Lichtenstein}},
  \bibinfo {author} {\bibfnamefont {A.~N.}\ \bibnamefont {Rubtsov}}, \bibinfo
  {author} {\bibfnamefont {M.}~\bibnamefont {Troyer}}, \ and\ \bibinfo {author}
  {\bibfnamefont {P.}~\bibnamefont {Werner}},\ }\href {\doibase
  10.1103/RevModPhys.83.349} {\bibfield  {journal} {\bibinfo  {journal} {Rev.
  Mod. Phys.}\ }\textbf {\bibinfo {volume} {83}},\ \bibinfo {pages} {349}
  (\bibinfo {year} {2011})}\BibitemShut {NoStop}%
\bibitem [{\citenamefont {Landau}\ and\ \citenamefont
  {Lifshitz}(1980)}]{landau1980statistical}%
  \BibitemOpen
  \bibfield  {author} {\bibinfo {author} {\bibfnamefont {L.~D.}\ \bibnamefont
  {Landau}}\ and\ \bibinfo {author} {\bibfnamefont {E.~M.}\ \bibnamefont
  {Lifshitz}},\ }\href@noop {} {\enquote {\bibinfo {title} {Statistical
  physics},}\ } (\bibinfo {year} {1980})\BibitemShut {NoStop}%
\bibitem [{\citenamefont {Sch\"ott}\ \emph {et~al.}(2016)\citenamefont
  {Sch\"ott}, \citenamefont {Locht}, \citenamefont {Lundin}, \citenamefont
  {Gr\aa{}n\"as}, \citenamefont {Eriksson},\ and\ \citenamefont
  {Di~Marco}}]{Schott15}%
  \BibitemOpen
  \bibfield  {author} {\bibinfo {author} {\bibfnamefont {J.}~\bibnamefont
  {Sch\"ott}}, \bibinfo {author} {\bibfnamefont {I.~L.~M.}\ \bibnamefont
  {Locht}}, \bibinfo {author} {\bibfnamefont {E.}~\bibnamefont {Lundin}},
  \bibinfo {author} {\bibfnamefont {O.}~\bibnamefont {Gr\aa{}n\"as}}, \bibinfo
  {author} {\bibfnamefont {O.}~\bibnamefont {Eriksson}}, \ and\ \bibinfo
  {author} {\bibfnamefont {I.}~\bibnamefont {Di~Marco}},\ }\href {\doibase
  10.1103/PhysRevB.93.075104} {\bibfield  {journal} {\bibinfo  {journal} {Phys.
  Rev. B}\ }\textbf {\bibinfo {volume} {93}},\ \bibinfo {pages} {075104}
  (\bibinfo {year} {2016})}\BibitemShut {NoStop}%
\bibitem [{\citenamefont {Pines}\ and\ \citenamefont
  {Nozi{\`e}res}(1966)}]{Pines66}%
  \BibitemOpen
  \bibfield  {author} {\bibinfo {author} {\bibfnamefont {D.}~\bibnamefont
  {Pines}}\ and\ \bibinfo {author} {\bibfnamefont {P.}~\bibnamefont
  {Nozi{\`e}res}},\ }\href@noop {} {\emph {\bibinfo {title} {The Theory of
  Quantum Liquids: Normal Fermi liquids}}}\ (\bibinfo  {publisher} {W.A.
  Benjamin, Philadelphia},\ \bibinfo {year} {1966})\BibitemShut {NoStop}%
\bibitem [{\citenamefont {Nordstr\"om}\ \emph {et~al.}(2016)\citenamefont
  {Nordstr\"om}, \citenamefont {Sch\"ott}, \citenamefont {Locht},\ and\
  \citenamefont {{Di Marco}}}]{GPU}%
  \BibitemOpen
  \bibfield  {author} {\bibinfo {author} {\bibfnamefont {J.}~\bibnamefont
  {Nordstr\"om}}, \bibinfo {author} {\bibfnamefont {J.}~\bibnamefont
  {Sch\"ott}}, \bibinfo {author} {\bibfnamefont {I.~L.~M.}\ \bibnamefont
  {Locht}}, \ and\ \bibinfo {author} {\bibfnamefont {I.}~\bibnamefont {{Di
  Marco}}},\ }\href@noop {} {\enquote {\bibinfo {title} {{A {GPU} code for
  analytic continuation through a sampling method}},}\ } (\bibinfo {year}
  {2016}),\ \bibinfo {note} {unpublished}\BibitemShut {NoStop}%
\end{thebibliography}

\end{document}